# Thermal imaging of dust hiding the black hole in the Active Galaxy NGC 1068

Violeta Gámez Rosas[1], Jacob W. Isbell[2], Walter Jaffe[1], Romain G. Petrov[3], James H. Leftley[3], Karl-Heinz Hofmann[4], Florentin Millour[3], Leonard Burtscher[1], Klaus Meisenheimer[2], Anthony Meilland[3], Laurens B.F.M. Waters[5,6], Bruno Lopez[3], Stéphane Lagarde[3], Gerd Weigelt[4], Philippe Berio[3], Fatme Allouche[3], Sylvie Robbe-Dubois[3], Pierre Cruzalèbes[3], Felix Bettonvil[7], Thomas Henning[2], Jean-Charles Augereau[8], Pierre Antonelli[3], Udo Beckmann[4], Roy van Boekel[2], Philippe Bendjoya[3], William C. Danchi[9], Carsten Dominik[10], Julien Drevon[3], Jack F. Gallimore[11], Uwe Graser[2], Matthias Heininger[4], Vincent Hocdé[3], Michiel Hogerheijde[1,10], Josef Hron[12], Caterina M.V. Impellizzeri[1], Lucia Klarmann[2], Elena Kokoulina[3], Lucas Labadie[13], Michael Lehmitz[2], Alexis Matter[3], Claudia Paladini[14], Eric Pantin[15], Jörg-Uwe Pott[2], Dieter Schertl[4], Anthony Soulain[16], Philippe Stee[3], Konrad Tristram[14], Jozsef Varga[1], Julien Woillez[17], Sebastian Wolf[18], Gideon Yoffe[2], Gerard Zins[14]

[1]Leiden Observatory, Leiden University, Leiden, Netherlands
[2]Max Planck Institute for Astronomy, Heidelberg, Germany
[3]Laboratoire Lagrange, Université Côte d'Azur, Observatoire de la Côte d'Azur, CNRS, Nice, France
[4]Max Planck Institute for Radio Astronomy, Bonn, Germany
[5] Department of Astrophysics, IMAPP, Radboud University, Nijmegen, Netherlands
[6] SRON Netherlands Institute for Space Research, Leiden, Netherlands
[7]ASTRON, Dwingeloo, Netherlands
[8]Université Grenoble Alpes, CNRS, IPAG, Grenoble, France
[9]Goddard Space Flight Center, NASA, Greenbelt Maryland, U.S.A.
[10]Anton Pannekoek Institute, University of Amsterdam, Amsterdam, Netherlands
[11]Dept. of Physics and Astronomy, Bucknell University, Lewisburg, Pennsylvania, U.S.A.
[12]Department of Astrophysics University of Vienna, Vienna, Austria
[13]Physikalisches Institut der Universität zu Köln, Köln, Germany
[14]European Southern Observatory, Santiago, Chile
[15]Centre d'Etudes de Saclay, Gif-sur-Yvette, France
[16]Sydney Institute for Astronomy, University of Sydney, Sydney, Australia
[17]European Southern Observatory, Garching bei München, Germany
[18]Institute of Theoretical Physics and Astrophysics, University of Kiel, Kiel, Germany

**Abstract**
**In the widely accepted 'Unified Model'[1] solution of the classification puzzle of Active Galactic Nuclei, the orientation of a dusty accretion torus around the central black hole dominates their appearance. In 'type-1' systems, the bright nucleus is visible at the centre of a face-on torus. In 'type-2' systems the thick, nearly edge-on torus hides the central engine. Later studies suggested evolutionary effects[2] and added dusty clumps and polar winds[3] but left the basic picture intact. However, recent high-resolution images[4] of the archetypal type-2 galaxy NGC 1068[5,6] suggested a more radical revision. They displayed a ring-like emission feature which the authors advocated to be hot dust surrounding the black hole at the radius where the radiation from the central engine evaporates the dust. That ring is too thin and too far tilted from edge-on to hide the central engine, and *ad hoc* foreground extinction is needed to explain the type-2 classification. These images quickly generated reinterpretations of the type 1-2 dichotomy[7,8]. Here we present new multi-band mid-infrared images of NGC1068 that detail the dust temperature distribution and reaffirm the original model. Combined with radio data [G, J. F. and I, C.M.V, in preparation], our maps locate the central engine below the previously reported ring and obscured by a thick, nearly edge-on disk, as predicted by the Unified Model. We also identify emission from polar flows and absorbing dust that is mineralogically distinct from that towards the Milky Way centre.**

# Main

## NGC 1068, the archetype Seyfert 2 galaxy

From the polarization properties of the Seyfert type 2 (Sy2) galaxy NGC 1068[6], Antonucci and Miller[1] established the Unified Model of Active Galactic Nuclei (AGNs), which explained the type 1-2 dichotomy with a physically and optically thick dust torus that hid the nucleus when viewed nearly edge-on. More recent observations and theory have refined the picture: the dust must be clumpy, its inner rim should be near the sublimation radius ($r_{sub}$), and radiation pressure can push the dust into an outflowing wind along the torus axis[9]. For NGC 1068, $r_{sub}$ is ~0.4 and ~0.2 parsecs (pc)[†] for silicate grains with radii between 0.005 and 0.25 μm, and for large graphite dust particles with radii>0.1 μm, which have sublimation temperature ($T_{sub}$) of ~1200 and ~2000 K, respectively, assuming a bolometric luminosity $L_{bol}=(0.4 - 4.7) \times 10^{45}$ erg/s (c.f. references [4,10,11] for details). Directly verifying the proposed torus geometry requires large interferometers to image these very small structures at infrared wavelengths (2-20 μm). Early MIDI/VLTI 10 μm interferometer data[12-14] was too limited to permit direct imaging, but model-fitting showed that complex, asymmetrically structured polar emission actually dominates the infrared luminosity. Recent extensive GRAVITY/VLTI data at 2 μm wavelength allowed a higher resolution image that showed an incomplete ring and other, fainter, structures[4]. The authors interpreted the ring as a planar circular feature of 0.24 pc radius seen 20° from edge-on and, after considering other possibilities, advocated identifying this feature as the hot sublimation ring with the central engine (CE) at its centre, primarily because its radius agreed with the expected value of $r_{sub}$ for large graphite grains. This geometry breaks with the Unified Model. The ring is too thin and too inclined to cover the CE and the authors invoke an additional component, a circumstantial foreground dust cloud to block the emission from the Broad Line Region (BLR). This break with the established model is based on the GRAVITY temperature estimate of the ring, and its consequent identification with $r_{sub}$, but in fact the wavelength range of GRAVITY is too restricted to establish accurate temperature values in the range of $T_{sub}$. The newly commissioned MATISSE/ESO/VLTI interferometer observes in the wavelength range from 3 to 13 μm, which is ideal for mapping the dust temperature distribution. NGC 1068 was therefore a major target for MATISSE observations.

## New infrared observations and images

In Sept. 2018 and Nov. 2019, we observed NGC 1068 with the MATISSE instrument (Methods §2, ref. 15) that combines very high angular resolution with a broad spectral range. The observations covered three infrared bands: *L* (3–4 μm), *M* (4.5–5 μm) and *N* (8–13 μm). The MATISSE interferometric data is densely enough sampled (Extended Data (ED) Figure 1) for image reconstruction. We utilized three independent image reconstruction systems: IRBis[16] MIRA[17], and a system based on the Högbom CLEAN algorithm[18] (Methods §6.6). They all converged to similar images. The IRBis images are shown in Figure 1.

In the *L*- and *M*- bands we see two similar structures: (1) a Northern asymmetric incomplete elliptical ring, 25x12 mas (1.7x0.9 pc) with Position Angle (PA)~-45°: the Northern Complex (NC), and (2) to the South, a fainter, less well-defined source: the Southern Extension (SE). The SE is not an interferometric artefact; it persists over wavelength and is independent of reconstruction method. The *N*-band images show a bright central source similar in extent and PA to the NC, and additional

---

[†] At distance of 14.4 Mpc, 1 pc subtends 14 mas.

diffuse structures at larger scales.

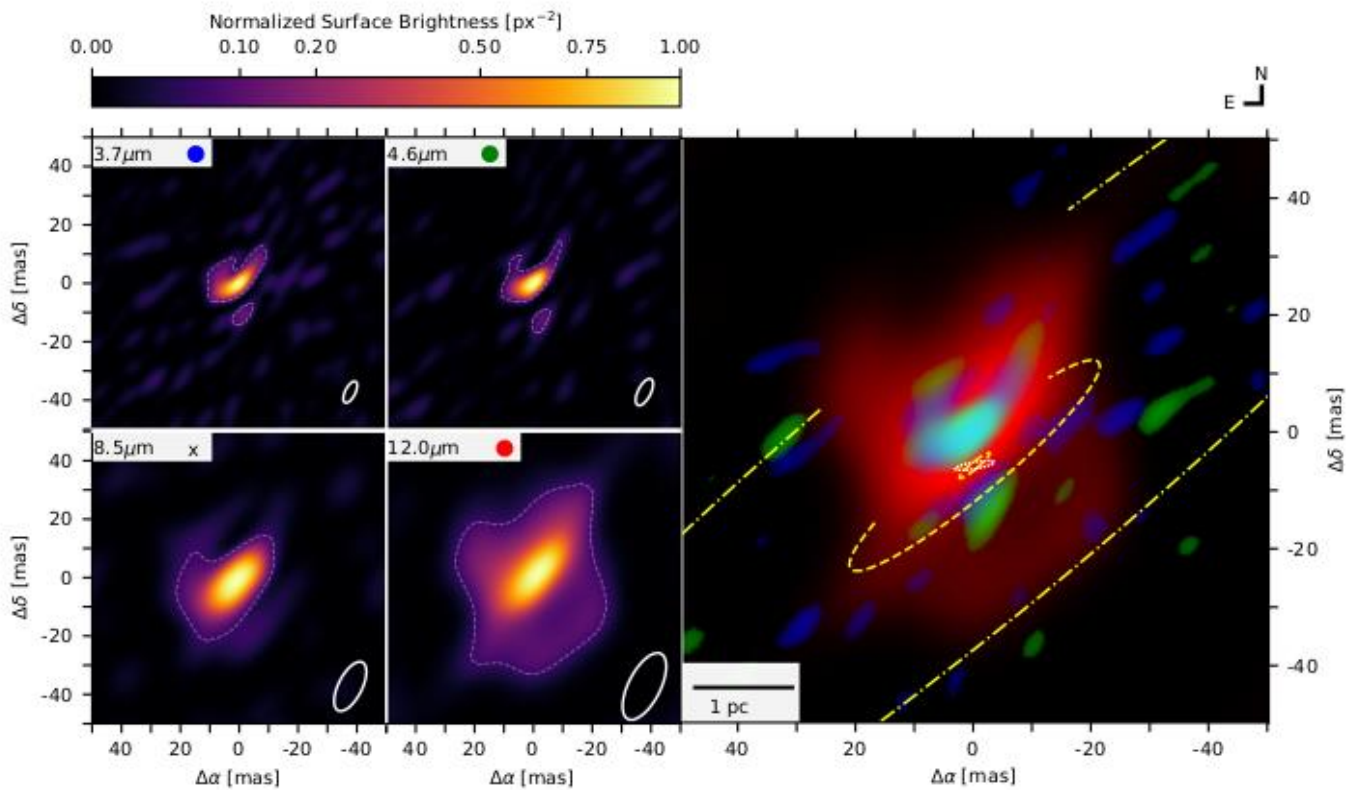

*Figure 1: IRBis reconstructed images of NGC 1068* in L-, M-, and N-bands at the labelled median wavelengths, plus an RGB-composite image of the 3.7, 4.6 and 12 μm images. The dashed white contours indicate the 3σ noise level, the white ellipses the resolution. The larger ellipses in the composite image illustrate the possible midplane of the obscuring layer discussed in the section on dust distribution; the outermost shows the maximum measured extent (10 pc). The innermost ellipses indicate two orientations of the hidden dust sublimation zone around the supermassive black hole, aligned with the ultraviolet polarization (white) and with the NC structures (yellow).

MATISSE data do not provide absolute positions that would allow direct registration of the different bands, so we have registered the images by two criteria: (1) morphological similarity; (2) consistency of the resulting spectral energy distributions (SEDs). This is equivalent to superimposing the brightest features in each band. This registration is shown in Figure 1. The SEDs extracted at various points after these registrations were smooth and physically plausible (c.f. Figures 2 and 3). This was not the case with significant (>3 mas) relative shifts of the images. Furthermore, we find a good match between NC and the *K*-band (2 μm) GRAVITY image[4] when we align the GRAVITY inner ring with the brightest component of NC (**Error! Reference source not found.**c). The *K*-band flux for E1 in Figure 3, agrees with our SED fit.

## Thermal modelling and dust mineralogy

To understand the thermal state of the dust in NGC 1068, we extracted spectrophotometry from nine areas or apertures based on the recovered geometry (Figure 2). We applied the apertures to the three reconstructed images and to a multi-Gaussian model fit of the data (c.f. Methods §6.5).

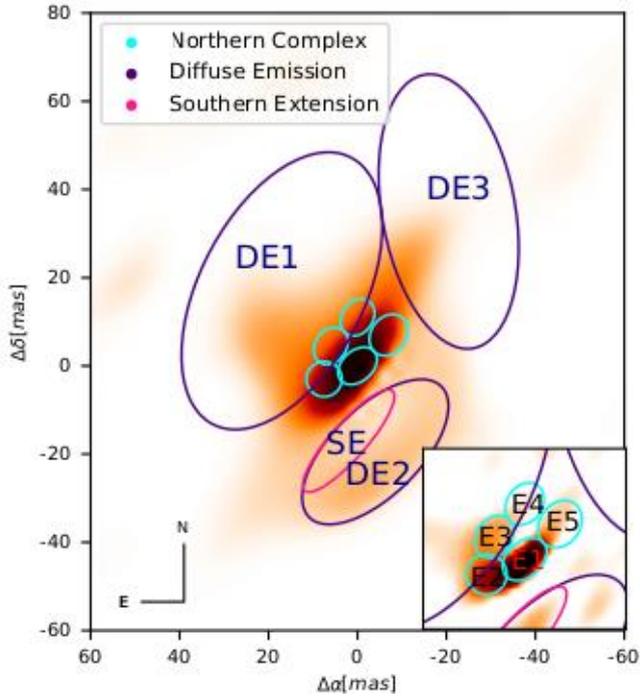

*Figure 2: Labelled large apertures for extraction of infrared SEDs,* superimposed on MATISSE N-band image. Inset: 30x30 mas enlargement of the central region (the Northern Complex) with labelled small apertures, superimposed on MATISSE L-band image.

We tested multiple image reconstructions and modelling methods and found that the fluxes extracted from each yielded equivalent results within the error bars (Methods §6.7.2, ED Figure 2). The Gaussian fits proved more convenient for computing SEDs at many wavelengths, as they provide uncertainty estimates via probability distributions, and are used below. The SEDs for the brightest central aperture E1 and for the high-temperature SE are plotted in Figure 3, the rest in ED Figure 3.

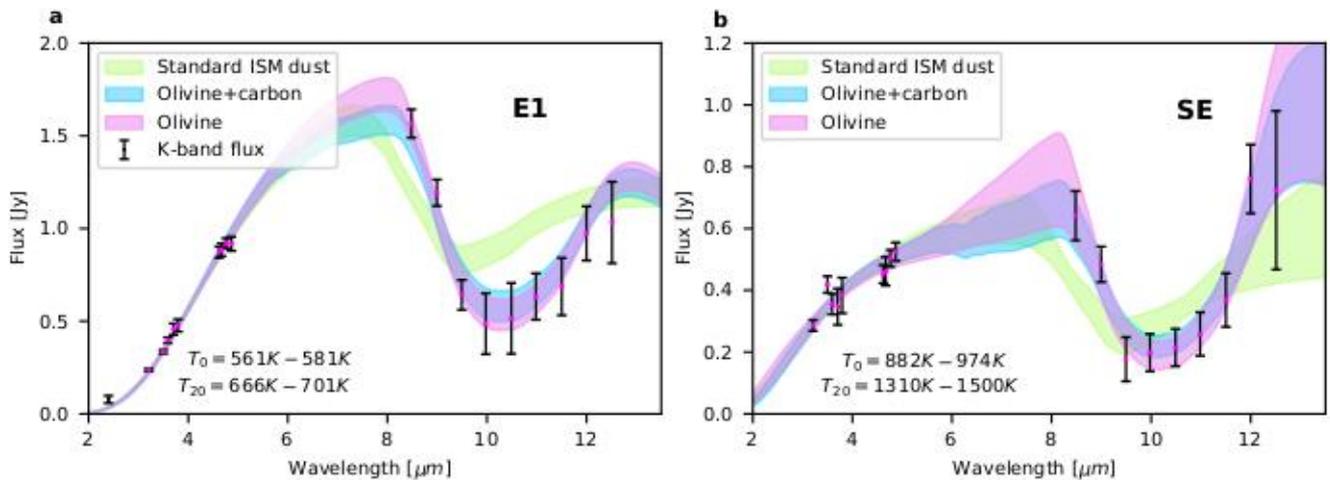

*Figure 3: Black body SED fits*: (a):1-temperature fit to the bright E1 component. The point at 2.2 μm gives the GRAVITY K-band flux (GRAVITY Collaboration priv. comm.) inside E1. (b): 2-Temperature blackbody fit to MATISSE SED of SE. The shaded regions confine models inside 1 sigma, colour-coded by different kinds of dust: green, cyan, and magenta correspond to Standard ISM dust, Olivine with 20% carbon and Olivine with no carbon, respectively. The colours are semi-transparent to show the regions where the models overlap. $T_0$ and $T_{20}$ refer to the hot component temperatures of the fits using 0% and 20% of carbon.

To add a physical interpretation to these SEDs, we modelled each with emission curves from a blackbody with a cold foreground absorbing screen. Most components required a second blackbody at a distinct temperature, a situation supported by radiation transfer modelling of dusty clumps[19]. We model each blackbody as:

$$S_\nu(\lambda) = A\eta\, B_\nu(T)\, \exp(-(\lambda)N_{ext})$$

$A$ is the area of each ellipse; $B_\nu(T)$ is the Planck function. The fitted parameters are the dust temperature $T$, the area filling factor $\eta$ for each blackbody, and the cold foreground dust surface density: $N_{ext}$ (μg cm$^{-2}$). $\kappa(\lambda)$ is the wavelength-dependent dust opacity per unit surface density. Contributions from silicates, iron, and carbon, as well as the dust grain size distribution determine $\kappa(\lambda)$ in the mid-infrared. The most prominent spectral feature is the strong silicate absorption near 10 μm. We first attempted fitting the SED of areas E1 and SE with a "standard interstellar" curve for $\kappa(\lambda)$ which closely resembles that measured toward the Galactic Centre[20,10]. This was unsuccessful; at 8 μm these curves contain a large opacity which is incompatible with the high measured fluxes, c.f. Figure 3. To clarify the dust composition, we tested theoretical opacity curves of common interstellar substances; we could only fit the measurements with dust particles comprised primarily of amorphous olivine ($[Mg_2,Fe_2]SiO_4$)[21] with an admixture of up to 20% by weight of carbonaceous dust and with relatively small dust particles (0.1–1.5 μm). These specifications apply to the foreground dust only; we have assumed the warm dust to emit as an optically thick blackbody. For the absorbing dust, $N_{ext}$ = 1000 μg cm$^{-2}$ corresponds to optical depths of $\tau_{3.4\mu m} = 2.0, \tau_{9.7\mu m} = 2.4, \tau_{10.6\mu m} = 2.5, \tau_{12\mu m} = 1.2$

The contribution of carbon is confirmed by the presence of a visibility feature close to 3.4 μm on several baselines, c.f. ED Figure 4, but no evidence of aromatic C-H features near 3.3 μm are seen. In preliminary high resolution N-band spectra (ED Figure 5) the shape of the silicate feature at the longest baselines suggests the presence of crystalline grains that would imply high temperature grain reprocessing. Further interpretation of these features will be presented in a future paper.

Figure 3 shows the best fits using pure olivine and a mix of olivine with carbon (cyan, magenta). Fits using standard ISM dust opacities (green) do not match the silicate absorption profile. The fraction by weight of carbon is limited to ≲20%; with the best fit near no carbon. Increased carbon fraction correlates with higher temperatures. We obtain a good fit with a single-black body for E1 and DE3, the other ellipses require 2 blackbodies. The temperature range for E1 is 560–700 K, while for SE the hot component requires temperatures from 880–1500 K. We present the map and table of temperatures for all the ellipses in Table 1.

| Aperture | Carbon content | $T_{cold}$ [K] | | | $T_{hot}$ [K] | | | $N_{ext}$ [μg/cm$^{-2}$] | | |
|---|---|---|---|---|---|---|---|---|---|---|
| | % | min | .. | max | min | .. | max | min | .. | max |
| E1 | 0 | 561 | .. | 581 | | N.A. | | 396 | .. | 547 |
| | 20 | 666 | .. | 701 | | | | 448 | .. | 610 |
| E2 | 0 | 188 | .. | 469 | 703 | .. | 1179 | 281 | .. | 530 |
| | 20 | 206 | .. | 506 | 829 | .. | 1461 | 386 | .. | 728 |
| E3 | 0 | 217 | .. | 376 | 860 | .. | 1095 | 67 | .. | 386 |
| | 20 | 225 | .. | 371 | 917 | .. | 1412 | 92 | .. | 489 |
| E4 | 0 | 232 | .. | 389 | 555 | .. | 818 | 281 | .. | 530 |
| | 20 | 250 | .. | 412 | 633 | .. | 1200 | 386 | .. | 621 |
| E5 | 0 | 223 | .. | 285 | 717 | .. | 1098 | 1 | .. | 100 |
| | 20 | 220 | .. | 287 | 712 | .. | 1275 | 1 | .. | 189 |
| DE1 | 0 | 229 | .. | 238 | 771 | .. | 865 | 1 | .. | 34 |
| | 20 | 228 | .. | 242 | 820 | .. | 870 | 1 | .. | 43 |
| DE2 | 0 | 139 | .. | 228 | 417 | .. | 475 | 825 | .. | 1000 |
| | 20 | 153 | .. | 292 | 543 | .. | 703 | 1000 | .. | 1468 |
| DE3 | 0 | 148 | .. | 162 | | N.A. | | 1 | .. | 197 |
| | 20 | 148 | .. | 162 | | | | 1 | .. | 241 |
| SE | 0 | 203 | .. | 354 | 846 | .. | 1031 | 530 | .. | 853 |
| | 20 | 173 | .. | 300 | 1333 | .. | 1500 | 464 | .. | 619 |

*Table 1: Blackbody model temperatures and extinction coefficients of the SEDs in each aperture. Next is in units of micro-grams cm-2. The SED fit for DE1 and DE2 uses the fluxes inside those apertures, without the contribution of the smaller overlapping apertures. The min. and max. temperatures and extinction values are defined as the range of values that produced fluxes within 1 sigma of the extracted SED. For the dust model used, a value of Next =1000 corresponds to τ=[2.0,2.4,2.5,1.2] at λ=[3.4,9.7,10.6,12] μm.*

For all apertures, $T_{hot}$ does not decrease systematically with distance from the inferred black hole position. For dust grains of fixed properties in radiation equilibrium from a central source we would expect $T \sim r^{-\alpha}$, with 0.3<α<0.5 depending on grain size. Although we only measure radii projected on the sky, the lack of a systematic temperature decrease with distance suggests heating by a mechanism other than direct radiation from the central engine.

The extinction varies greatly on small scales, implying it arises close to the sources. It increases rapidly and systematically to the South-West of E3, extending at least 30 mas in projection from E1 to cover SE and DE2 with the highest opacities. The Northern diffuse apertures show almost no extinction.

## Radio images and the hidden black hole

In Figure 4 we compare MATISSE images to an ALMA continuum image at 1.2 mm wavelength[22] and a VLBA image at 1.4 cm (Methods §7, G, J. F. and I, C.M.V, in preparation) with the positions of a series of water masers[23] which mark warm, dense molecular gas. The morphological similarities between infrared and radio images are striking. The smoothed MATISSE 12 µm image and the ALMA image show X-shaped emission of similar dimensions and orientation which can be registered to <3 mas accuracy (Methods §7.1). With this cross-identification, the 3.7 µm MATISSE image and the VLBA image can be superimposed (Figure 4b) to the same accuracy (*idem*). The strong resemblance in shape and orientation of the Northern arc of the VLBA and the *LM*-band NC supports this alignment. This fixes the position of the 22 GHz peak, which we assume marks the position of the black hole, at slightly below the lower edge of component E2, and sets the positions of the masers along the lower edge of NC.

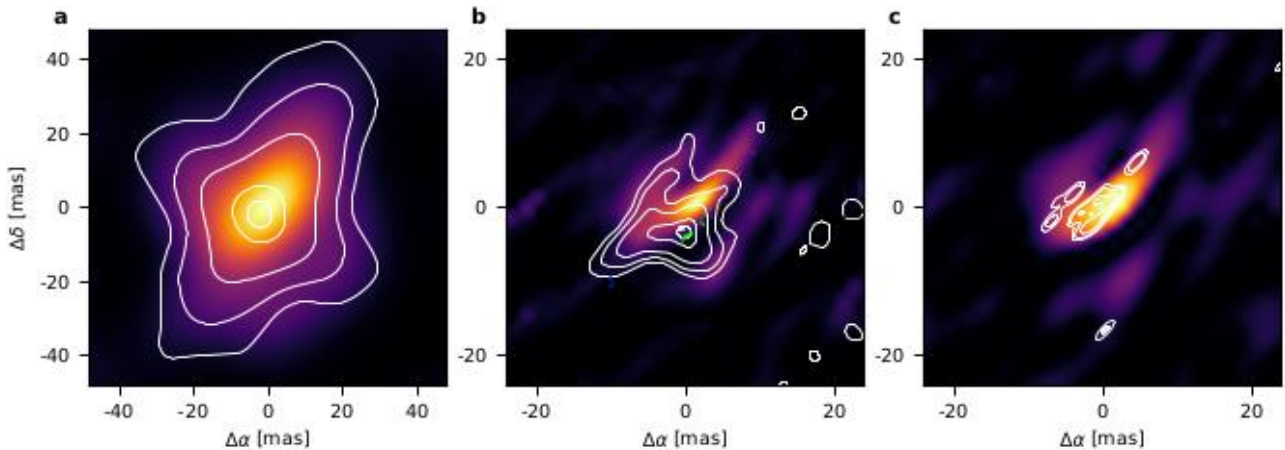

*Figure 4: Comparison of infrared and radio images:* (a) MATISSE N-band image, and contours of ALMA 256 GHz radio image. The contours are [0.1, 0.2,0.4,0.8,0.95] of peak (0.85 mJy/beam). (b) MATISSE L-band image and contours of VLBA 22 GHz image. The white asterisk marks the probable black hole position. The small dots mark the maser positions and their colours indicate their redshifts relative to the galaxy systemic velocity. Note the smaller field of view. The contour levels are [0.1,0.2,0.4,0.8,0.95] of the peak (3.0 mJy/beam) (c): MATISSE L-band image and contours of GRAVITY K-band image. Contours are [0.1,0.2,0.5] of the peak.

The centimetre and millimetre radio emission arise as bremsstrahlung from very hot ionized gas ($>\sim 10^6$ K)[24] (discussed in Methods §7). The association between this gas and the warm dust structures on 1-6 pc scales is puzzling because, if co-spatial, the gas would destroy the dust on short time scales. The association may only be a projection on the plane of the sky, but the morphological similarity strongly suggests connected physical processes. Both sources resolve into complex morphologies that cannot be characterized by a single PA. The dominant axis of the millimetre source seems to be near PA 30° and that of the centimetre source is near -60°, although considering the axis of the extensions from the NW to the SE we obtain PA~60° and -30°, which is

## Spatial distribution of the dust

We estimate the temperature of the brightest emission feature, E1, to be 684±17 K, well below $T_{sub}$. Together with the position of the VLBA peak and the gradients in temperature and extinction toward the South, this strongly suggests that the black hole and sublimation rim lie below E1 and E2 and are hidden in *LM*-bands by a band of highly optically thick dust, which is responsible for the observed type 2 classification. The opacity weakens to the South to reveal the SE. The proposed position of the black hole and temperatures we recover agree better with the interpretation of the brightest feature of the GRAVITY image (Figure 4c) as cool (720 K) dust (their model 2), rather than a hot disk[4]. A layer of cold dust with $\tau_{12\mu}$~0.8 covers the whole SW quadrant, including DE2, to a projected distance of ~2.5 pc, but does not cover DE1 or DE3 in the North. The overall pattern is consistent with a large sheet or disk of dust, inclined away from us, with major axis at PA ~-45°, that covers the Southern structures, but exposes DE1 and DE3. The disk opacity is highest in a 0.3 pc region between E1/E2 and SE, covering the central engine, but it flares up to partly cover the NC. On scales of 50-100 pc, the geometry of the Narrow Line Region and other structures implies that its pole is about 10° from the plane of the sky[25,26], a similar inclination to the line of sight would imply a radial extent of our obscuring sheet in excess of 10 pc. The extent and orientation of this obscuring sheet at this inclination are indicated in Figure 1. The components DE1 and DE2 probably represent dust blown out of the centre; their orientation matches wind structures at larger scales[27]. The winds may originate within the hot irregular NC and SE structures.

In conclusion, we find the dusty structure that plays the role of the "historical torus" with an optically thick ring obscuring the central engine at parsec scales and a less optically thick disk extending to at least 10 pc. The distribution of the warm and hot dust reveals two main axes, one along the densest obscuring structure, at position angle -45°, and a second perpendicular axis parallel to the molecular outflows. However, the detailed structures are asymmetric in all bands, suggesting that they are not in full thermal and dynamical equilibrium. This conclusion is supported by the difference in PA between the minor axis of NC (~45°) which represents the angular momentum of the dust system at pc scales, and the PA of the radio jets (~10°)[28,29], which indicates the black hole spin axis, and the PA perpendicular to the ultraviolet polarization directions (~7°)[30,31], which may indicate the spin axis of the sublimation ring. The innermost ellipses in Fig. 1 show two possibilities for the orientation of this ring: aligned with the ultraviolet polarization (white) and with the NC structures (yellow). The cold dust grain properties appear quite different from the standard Milky Way ISM dust. The grains are sub-micron Mg-rich olivines with some carbon. The corresponding properties of the hot dust seen in emission remain open issues. Future MATISSE observations of additional AGNs will show whether NGC 1068 is atypical or prototypical. These high resolution, broadband images from MATISSE open new avenues for investigations of the dust-radiation interactions in AGNs, such as detailed radiation transfer modelling of the clouds, studies of the evolution of carbon and silicate grain mineralogy in AGN environments, and more globally, a synthesis of observations with radio, millimetre, and infrared instruments to connect the dust cloud physics on scales of tenths to tens of parsecs.

## References

1. Antonucci, R. Unified models for active galactic nuclei and quasars. *Ann. Rev. Astron. Astrophys.* **31**, 473–521 (1993).
2. López-Gonzaga, N. & Jaffe, W. Mid-infrared interferometry of Seyfert galaxies: Challenging the Standard Model. *Astron. Astrophys.* **591**, A128 (2016).
3. Asmus, D.; Hönig, S. F.; Gandhi, P., The Subarcsecond Mid-infrared View of Local Active Galactic Nuclei. III. Polar Dust Emission, *Astrophys. J.,* **822,** 109-121 (2016).
4. GRAVITY Collaboration. An Image of the Dust Sublimation Region in the Nucleus of NGC 1068. *Astron. Astrophys.,* **634,** 1 (2020).
5. Seyfert, Carl K., Nuclear Emission in Spiral Nebulae. *Astrophys. J.* **97,** 28-40 (1943).
6. Antonucci, R. R. J. & Miller, J. S. Spectropolarimetry and the nature of NGC 1068. *Astrophys. J.* **297**, 621–632 (1985).
7. Vermot, P. *et al.* The hot dust in the heart of NGC 1068's torus: A 3D radiative model constrained with GRAVITY/VLTI. *arXiv e-prints* arXiv:2106.04211 (2021).
8. Prieto, A., Nadolny, J., Fernández-Ontiveros, J. A. & Mezcua, M. Dust in the central parsecs of unobscured AGN: more challenges to the torus. *Mon. Not. R. Astron. Soc.* **506**, 562-580 (2020)

9. Hönig, S. F. Redefining the Torus: A Unifying View of AGNs in the Infrared and Submillimetre. *Astrophys. J.* **884**, 171 (2019).
10. Barvainis, R. Hot dust and the near-infrared bump in the continuum spectra of quasars and active galactic nuclei. *Astrophys. J.* **320**, 537–544 (1987).
11. Baskin, A. & Laor, A. Dust inflated accretion disc as the origin of the broad line region in active galactic nuclei. *Mon. Not. R. Astron. Soc.* **474**, 1970–1994 (2018)
12. Jaffe, W. et *al.* The central dusty torus in the active nucleus of NGC 1068. *Nature* **429**, 47–49 (2004).
13. Raban, D., Jaffe, W., Röttgering, H., Meisenheimer, K. & Tristram, K. R. W. Resolving the obscuring torus in NGC 1068 with the power of infrared interferometry: revealing the inner funnel of dust. *Mon. Not. R. Astron. Soc.* **394**, 1325–1337 (2009).
14. López-Gonzaga, N., Burtscher, L., Tristram, K. R. W., Meisenheimer, K. & Schartmann, M. Mid-infrared interferometry of 23 AGN tori: On the significance of polar-elongated emission. *Astron. Astrophys.* **591**, A47-65(2016).
15. Lopez, B. et al. MATISSE, the VLTI mid-infrared imaging spectro-interferometer. *arXiv e-prints* arXiv:2110.15556 (2021)
16. Hofmann, K.-H., Weigelt, G. & Schertl, D. An image reconstruction method (IRBis) for optical/infrared interferometry. *Astron. Astrophys.* **565**, A48 (2014).
17. Thiébaut, E. MIRA: an effective imaging algorithm for optical interferometry. in *Optical and Infrared Interferometry* vol. **7013** 479–490 (SPIE, 2008).
18. Högbom, J. A. Aperture Synthesis with a Non-Regular Distribution of Interferometer Baselines. *Astron. Astrophys. Suppl. Series* **15**, 417 (1974).
19. Nenkova, M., Sirocky, M. M., Ivezic ́, Z, Elitzur, M., AGN Dusty Tori. I. Handling of Clumpy Media. *Astrophys. J.* **685**, 147-159 (2008)
20. Fritz, T. K. *et al.* Line Derived Infrared Extinction toward the Galactic Center. *Astrophys. J.* **737**, 73 (2011)
21. Min, M., Hovenier, J. W. & de Koter, A. Modelling optical properties of cosmic dust grains using a distribution of hollow spheres. *Astron. Astrophys.* **432**, 909–920 (2005).
22. Impellizzeri, C. M. V. *et al.* Counter-rotation and High-velocity Outflow in the Parsec-scale Molecular Torus of NGC 1068. *Astrophys. J. Letters* **884**, L28-L33 (2019).
23. Greenhill, L. J., Gwinn, C. R., Antonucci, R. & Barvainis, R. VLBI Imaging of Water Maser Emission from the Nuclear Torus of NGC 1068. *Astrophys. J. Letters* **472**, L21-L24 (1996).
24. Gallimore, J. F., Baum, S. A. & O'Dea, C. P. The Parsec-Scale Radio Structure of NGC 1068 and the Nature of the Nuclear Radio Source. *Astrophys. J.* **613**, 794–810 (2004)
25. Das, V., Crenshaw, D. M., Kraemer, S. B. & Deo, R. P. Kinematics of the Narrow-Line Region in the Seyfert 2 Galaxy NGC 1068: Dynamical Effects of the Radio Jet. *Astron. J.* **132**, 620–632 (2006).
26. Poncelet, A., Sol, H. & Perrin, G. Dynamics of the ionization bicone of NGC 1068 probed in mid-infrared with VISIR. *Astron. Astrophys.* **481**, 305–317 (2008).
27. Garcia-Burillo, S. *et al.*, ALMA images the many faces of the NGC 1068 torus and its surroundings, *Astron. Astrophys.* **632**, A61 (2019)
28. Evans, I. N. et al. HST Imaging of the Inner 3 Arcseconds of NGC 1068 in the Light of [O III] lambda 5007. *Astrophys. J.* **369**, L27 (1991).
29. Gallimore, J. F., Baum, S. A., O'Dea, C. P. & Pedlar, A. The Subarcsecond Radio Structure in NGC 1068. I. Observations and Results. *Astrophys. J.* **458**, 136 (1996).
30. Antonucci, R., Hurt, T. & Miller, J. HST Ultraviolet Spectropolarimetry of NGC 1068. *Astrophys. J.* **430**, 210 (1994).
31. Kishimoto, M. The Location of the Nucleus of NGC 1068 and the Three-dimensional Structure of Its Nuclear Region. *Astrophys. J.* **518**, 676–692 (1999).

## Main Figure Legends:

**Figure 1: IRBis reconstructed images of NGC 1068** in L-, M-, and N-bands at the labelled median wavelengths, plus an RGB-composite image of the 3.7, 4.6 and 12 μm images. The dashed white

contours indicate the 3σ noise level, the white ellipses the resolution. The larger ellipses in the composite image illustrate the possible midplane of the obscuring layer discussed in the section on dust distribution; the outermost shows the maximum measured extent (10 pc). The innermost ellipses indicate two orientations of the hidden dust sublimation zone around the supermassive black hole, aligned with the ultraviolet polarization (white) and with the NC structures (yellow).

**Figure 2: Labelled large apertures for extraction of infrared SEDs,** superimposed on MATISSE N-band image. Inset: 30x30 mas enlargement of the central region (the Northern Complex) with labelled small apertures, superimposed on MATISSE L-band image.

**Table 1: Blackbody model temperatures and extinction coefficients of the SEDs in each aperture**. Next is in units of micro-grams cm-2. The SED fit for DE1 and DE2 uses the fluxes inside those apertures, without the contribution of the smaller overlapping apertures. The min. and max. temperatures and extinction values are defined as the range of values that produced fluxes within 1 sigma of the extracted SED. For the dust model used, a value of Next =1000 corresponds to τ=[2.0,2.4,2.5,1.2] at λ=[3.4,9.7,10.6,12] μm.

**Figure 3: Black body SED fits**: (a):1-temperature fit to the bright E1 component. The point at 2.2 μm gives the GRAVITY K-band flux (GRAVITY Collaboration priv. comm.) inside E1. (b): 2-Temperature blackbody fit to MATISSE SED of SE. The shaded regions confine models inside 1 sigma, colour-coded by different kinds of dust: green, cyan, and magenta correspond to Standard ISM dust, Olivine with 20% carbon and Olivine with no carbon, respectively. The colours are semi-transparent to show the regions where the models overlap. $T_0$ and $T_{20}$ refer to the hot component temperatures of the fits using 0% and 20% of carbon.

**Figure 4: Comparison of infrared and radio images:** (a) MATISSE N-band image, and contours of ALMA 256 GHz radio image. The contours are [0.1, 0.2,0.4,0.8,0.95] of peak (0.85 mJy/beam). (b) MATISSE L-band image and contours of VLBA 22 GHz image. The white asterisk marks the probable black hole position. The small dots mark the maser positions and their colours indicate their redshifts relative to the galaxy systemic velocity. Note the smaller field of view. The contour levels are [0.1,0.2,0.4,0.8,0.95] of the peak (3.0 mJy/beam) (c): MATISSE L-band image and contours of GRAVITY K-band image. Contours are [0.1,0.2,0.5] of the peak.

# Methods

## 1 Introduction

Here we detail the MATISSE (Multi AperTure mid-Infrared Spectro-Scopic Experiment)[15] instrument, the observations, and the data reduction. We compare the results of several different image reconstruction techniques for our data. We describe the methods for extracting spectrophotometric data from several spatial areas of interest, and for fitting multi-component black-body emission models to these spectra, as presented in the main paper. Finally, we present details of the radio observations and the results of the black-body fits for those apertures not given in the main paper.

## 2 Description of Instrument

MATISSE is a mid-infrared spectro-interferometer that combines the beams of four telescopes. For more details see the MATISSE reference paper[15]. MATISSE provides visibilities for six telescope pairs and closure phases for each combination of three telescopes. The closure phases contain information about the geometry of the sources and their distribution on the sky that is free from atmospheric phase errors.

MATISSE introduces to interferometry the *L*- and *M*-bands (3.0—4.0 μm and 4.6—5.0 μm, respectively); and includes the *N*-band (8.0—13.0 μm) previously observed with MIDI[32] (MID-infrared Interferometer), thus opening a new window for a more complete study of AGNs. The *LM*-band and *N*-band subsystems operate simultaneously but in separate cryostats, with separate cold optics.

MATISSE operates at ESO's VLTI in Cerro Paranal (Chile), and uses either the four 8.2 m Unit Telescopes (UTs) in their fixed positions, or the four relocatable 1.8 m Auxiliary Telescopes (ATs) in various configurations. This means that the baselines span lengths from about 15 to 140 metres, resulting in angular resolutions as high as 3 mas in the *L*-band and 10 milliarcseconds (mas) in the *N*-band. The design considerations, detailed performance, and operational characteristics of MATISSE are described in the MATISSE reference paper[15].

## 3 Description of the observations

NGC 1068 was observed during MATISSE commissioning in September 2018 (under program 0103.C-0143) and again with Guaranteed Time Observation (GTO) in 2019 (under 0104.B-0322). We use both datasets in our work, both are available at the ESO archive. The detailed logs of both the science and calibrator observations are in the site described in the Data Access section, and the *uv*-coverage is shown in Extended Data Figure 1.

The instrument was configured in spectrally resolved Low Resolution "Hybrid" mode. In this mode the *LM*-band optical system uses a prism dispersing element (R≃30) in the Simultaneous Photometry mode (SIPHOT), where a 1-to-3 beam-splitter separates the photometric and interferometric signals which are detected simultaneously. This allows *visibilities*, the ratio of these signals, to be measured in a way that is largely insensitive to the variations in the atmospheric Strehl ratio, which are important at these wavelengths. The *N*-band observations use an R~30 prism in High Sensitivity (HISENS) mode, because the high sky backgrounds require separate photometric observations, and the Strehl ratio variations are less important. A detailed description of the available MATISSE observing modes can be found in the MATISSE reference article and at the ESO website[‡].

To calibrate the instrument and atmosphere, we chose *interferometric* calibrator stars from a catalogue of mid-IR interferometric standard stars[33], that were at similar airmass as the target (<10° from the target), within a factor 2 in either *L*- or *N*- flux to the target, and whose angular diameters are small and well known.

---

[‡] https://www.eso.org/sci/facilities/paranal/instruments/matisse.html

The calibrators for each observation are given in Observing log (c.d. Data Access) along with the quantity $\tau_o$ which describes the atmospheric coherence time in the *V*-band in milliseconds; it is a measure of the reliability of the visibilities. It has proved to be a more consistent estimator of data quality than measurements of the atmospheric seeing. Typically, $\tau_o$ >5 ms indicates high reliability, while $\tau_o$ <3 ms indicates low reliability[15]. Most of the commissioning data included here were taken under intermediate (~3–5 msec) to good (5+ msec) conditions; the GTO data were mostly taken under good conditions.

A standard observation contains six visibilities, from the distinct pairings of the four telescopes. The projections of the baseline vectors on the plane of the sky at the source are the *uv*-coordinates of the observations, and these coordinates specify which components of the Fourier transform of the sky brightness are represented in the visibilities. Extended Data Figure 1 shows the combined *uv*-coordinates of all our measurements. The maximum extent of the *uv*-coverage sets the effective resolution of the interferometric synthetic images; incompleteness in the filling of the *uv*-plane tends to create "aliased" artefacts in these images.

# 4 Data flagging, reduction and calibration

## 4.1 Introduction

In this section we describe data measured in the *L*-, *M*-, and *N*-bands during the September 2018 commissioning runs and the GTO observation of November 2019. Typical of commissioning, not all data were of useful quality, either because of instrumental problems or because of bad weather, and we have excluded data deemed to be less reliable. In some cases, we applied special reduction steps that are not necessary for routine operations.

In the following sections we describe:

- Pre-reduction data selection and editing
- Pipeline reduction of raw data
- Post-reduction selection and editing

## 4.2 Data selection and flagging

### 4.2.1 *LM*-band chopping status

When estimating non-interferometric photometry, we only include exposures during which the telescope secondary mirrors were chopped at ~1 Hz between target and sky to allow accurate removal of sky backgrounds.

### 4.2.2 Revision of chopping status labels

The MATISSE operating software contains problems that sometimes resulted in mislabelled chopping status of individual frames as T/U/S=Target/Unknown/Sky in the archived data files. In this project we corrected this with specially written software. Unfortunately, this labelling problem has persisted but the current versions of the MATISSE pipeline procedures detect and correct it automatically.

## 4.3 Pipeline Data reduction

We removed known instrumental signatures from the remaining observations and extracted interferometric measures with the procedures described below. We then selected reliable exposures based on signal/noise (S/N) criteria, and finally calibrated the reduced measures using the calibrator observations. The output measures are the squared visibilities for each baseline, the closure phases for each triplet of baselines, and the photometric (one-telescope) flux, averaged over the set of exposures. Each measure is a function of wavelength.

We reduced the data with the ESO/MATISSE ESOREX Data Reduction Software (DRS) pipeline. The DRS evolved during our analysis to accommodate newly discovered issues, and different versions were applied to different data subsets. For most *LM*-band data we used version 1.4.1; for *N*-band data and "binned" *LM*-band data (c.f. §4.3.1) we used version 1.5.2 which contained new algorithms specific to these modes. As of version 1.5.2 of the DRS, the phase convention of the *N*-band is flipped with respect to the phase convention of the *L*- and *M*-bands. This causes a 180° rotation in the final images (see §4.8). The data and images of this paper have been corrected for this.

Reduction of ground-based interferometric data usually follows one of two strategies:

- **Coherent processing:** wherein estimates of atmospheric phase effects are made and removed before averaging data in time. The typical output measures are *coherent fluxes* and *chromatic phases*.
- **Incoherent processing:** wherein the previous approach is ineffective because of rapid variations of phase. Output measures are chosen that are insensitive to these variations, at some cost in final Signal/Noise ratio for faint sources. These outputs are *squared visibilities* and closure *phases*. For computing visibilities, single-telescope photometric data are also taken, generally with telescope chopping to measure and remove the sky and telescope backgrounds

Generally, incoherent methods are used at shorter wavelengths where atmospheric phases vary rapidly, and coherent methods at longer wavelengths where detector and thermal noise issues dominate. The mid-infrared is typically the transition region between these two methods. Here we have reduced data with the incoherent method in contrast to the previous 10 μm MIDI data that was reduced in coherent mode[34], because at the current time this method has been better validated. In the detailed context of the MATISSE DRS, we specified this procedure by using the parameters corrFlux=FALSE, OpdMod=FALSE and compensate=[pb,nl,if,rb,bp,od] for the master recipe mat_raw_estimates. The reduced squared visibilities (**vis\*\*2**) and closure phases are shown in ED Figures 4 and 6.

### 4.3.1 Spectral Binning

Extracting closure phases requires taking the triple product of the three baselines included in a "triangle" of three telescopes. This non-linear procedure greatly magnifies the effect of noise if the noise per frame and per spectral channel is <<1. We mitigate this effect by binning the raw data in the spectral direction before extracting the triple products. This binning can however reduce the visibility contrast when the interferometric phase varies rapidly with wavelength. In the low-resolution mode that we used, both detectors are oversampled, and a modest binning increases the Signal/Noise ratio while avoiding contrast loss. For *LM*-band we use the default DRS bin size of 5 pixels (0.25 μm) for the visibilities and closure phases; in the *N*-band we use 7 pixels (0.3 μm) for the visibilities. We use 21 pixels (1 μm) for the closure phases in the *N*-band. However, at the shortest *L*-band wavelength (3.2 μm), the low single exposure S/N yielded poor closure phase accuracy. We separately reduced the data at 3.2 μm using a DRS binning parameter ("--spectral binning") of 11 pixels (0.5 μm) for closure phases, retaining the original reduced data at 5-pixel binning for the visibilities. We find that over the entire band, closure phase uncertainties were decreased from ~10° to ~3° on as a result of binning.

### 4.4 Data selection

After the pipeline reduction, we selected reliable sub-exposures by evaluating the frame-to-frame stability of both the photometric and interferometric data. Low photometric fluxes generally indicate a failure of the VLTI Adaptive Optics (AO) system to correct the atmospheric seeing. Low interferometric fluxes can be caused by either AO failure or extremely unstable atmospheric conditions. The versions of DRS that we use does not support this frame-by-frame inspection; therefore, we used the EWS package[35] adapted from the MIDI reduction system. Future versions of DRS will include this statistical grading option. The final post-processing of the accepted data sets was performed in DRS.

Using EWS, we flagged a frame as acceptable if it met selection criteria based on its associated photometry and correlated flux.

A photometry frame was rejected if the photometry counts in a frame were less than either:

- 0.5 times the 80th percentile of counts for the single-dish telescope flux
- 6 times the RMS noise for all frames

For interferometric frames, we first estimated the correlated flux as the peak of the Fourier transform of the data as a function of wave number. We reject frames in which this peak is less than either:

- 0.001 times the 85th percentile of all baselines and frames (to eliminate weak "ghosts" caused by internal reflections)
- 0.5 times the 85th percentile of fluxes for this baseline or less than 7 times the rms noise (*LM*-bands) or 9 times the noise (*N*-band). This eliminates frames where the atmosphere was exceptionally unstable or the AO system failed.

Any sub-exposure in which more than 50% of the frames passed both criteria was accepted for DRS processing. In general, either just a *few* or *almost all* frames of a given sub-exposure are adversely affected by AO or atmospheric problems. The above criteria gave a clean separation between these two cases.

## 4.5 Calibration of the target measures

For the calibration of the interferometric data of NGC 1068 we used the esorex recipe mat_cal_oifits, including only the selected sub-exposures of the target and calibrators. The different BCD settings that we used have different instrumental signatures, and are calibrated separately. We then averaged together the calibrated visibilities and closure phases for the various settings. An in-depth description of the BCD calibration and the sign convention can be found in the detailed AMBER data reduction[36]. The final calibrated squared visibilities and closure phases are plotted in Extended Data Figure 4 and Extended Data Figure 6.

## 4.6 Photometric calibration

While MATISSE is an interferometric instrument, in SIPHOT mode, it also captures a *single dish* spectrum from each individual telescope. For an extended target, such as NGC 1068, this represents the flux through a circular aperture of diameter 0.135". This low-resolution spectrum has 64 spectral channels between 2.9 $\mu$m and 5.0 $\mu$m (the *L*- and *M*-bands). To connect these spectra to an absolute flux scale, we observed two standard stars with known model-atmosphere spectra from the Cohen catalogue[37]. These stars were HD 9362 (del Phe), observed on both observing nights, and HD 188603 (b Sgr), on September 25th only. These stars are too well-resolved to serve as interferometric calibrators. In the N-band we followed a similar procedure, using the Cohen calibrator HD 18322 (Eta Eri) as standard. On each night the photometric spectra from the interferometric calibrator stars were cross-calibrated with the Cohen standard stars to bring them to an absolute scale. These spectra were then applied to the instrumental photometric spectra of NGC 1068 to bring them to the same scale. All of the spectra so calibrated were averaged together to give the final target spectrum, and the errors in the process were estimated by variations between the individual spectra. The final photometric spectrum of NGC 1068 is shown in Extended Data Figure 5, with comparison to broad-band photometric estimates from other instruments.

## 4.7 Faint-source calibration

Although NGC 1068 is the mid-infrared brightest AGN in the southern sky, it is still a relatively faint source for MATISSE. In order to verify that the data processing software produces reliable results for faint sources, we performed a number of tests on calibrators. This was particularly important as the analysis of NGC 1068 data started early during the commissioning of MATISSE. Our specific tests on that target were eventually confirmed and strengthened by the global commissioning of MATISSE[38].

It was particularly important to investigate the linearity of MATISSE measures at low single-dish flux and low coherent flux since NGC 1068 is a faint and extended source. To check the linearity of MATISSE measures at low flux, we used a large set of calibrators[§] with decreasing fluxes. As in the NGC 1068 measurements, these observations were made with the UTs in LOW resolution. In L and M bands we used the SIPHOT mode (photometric and interferometric channels recorded simultaneously) on observations chopped at 0.5 Hz. In N we used the HIGHSENS mode with chopped photometric observations after the non-chopped interferometric observations on the same detector pixels. The angular diameters of the calibrators were taken from the JMMC Stellar Diameter Catalogue[33,39]. Errors on the diameters of these faint and hence quite unresolved targets convert into measured visibility errors much smaller than 1%, i.e., well under the fundamental noise errors. From the calibrator observations, we generated transfer functions of the Instrumental Squared Visibility (ISV) and the closure phases (T3PHI). We compare these to flux, $\tau_o$, and time (the observations were made in different dates over the three different months), at the selected wavelengths (3.4 µm, 3.6 µm, 3.8 µm in L; 4.7 µm in M and 8.5 µm, 11 µm and 12 µm in the N band). The ISV is defined as the squared visibility (squared correlated flux/photometric flux), corrected for the known finite source sizes. In all cases we found a linear and stable ISV against flux, as well as for the closure phases against flux, as illustrated by Extended Data (ED) Figure 1. Eventually, the general commissioning of MATISSE set the following limits for MATISSE unbiased observations and visibility error<0.1 and closure phase error<5° per spectral channel and 1 min exposure:

- For the L band, flux>0.3 Jy and coherent flux >0.06 Jy
- For the N band, flux>0.9 Jy and coherent flux >0.4 Jy

For NGC 1068 we have a flux ~1 Jy at 3 µm and ~2 Jy at 4 µm and >7 Jy in N. This is always quite far above the limit for photometric calibration. The coherence flux limit then sets an unbiased visibility limit as low as $V^2$~0.008 in L and $V^2$~0.002. All the measured values (except for the data sets discarded by the selection process for technical or seeing problems) are above these limits as shown in ED Figures 4 and 6. The single exception might be some very low visibility points on long baselines, but then the measurements and their bias are well below the fundamental errors. These points are considered to be zero within noise.

This underlines our confidence that both MATISSE and the DRS deliver reliable measures for faint sources with fluxes comparable to NGC 1068.

## 4.8 Determination of Image Orientation

In complicated interferometric systems, there is sometimes at commissioning time an uncertain phase sign ambiguity which translates to a 180° rotation ambiguity in reconstructed images. To validate the proper orientation of each band, we both performed image reconstruction and fit two Gaussian sources to the known binary δ Sco. This unequal binary was observed on 19 May 2018

---

[§] We have ~50 measurements (i.e., ~12 targets) fainter than 20 Jy with ATs in L and in N which corresponds to the range <1 Jy with UTs). This was confirmed by a few targets <0.1 Jy in L on UTs and <0.5 Jy in N)

during MATISSE commissioning. Based on previous data[40], we expect the separation at this time to be 178 mas with the fainter star almost directly to the North of the brighter star. In the model fitting, however, we do not enforce this assumption, and instead use uniform priors over the parameter space. Both image reconstructions and Gaussian modelling of this source reproduce the angular separation of the binary. From these reconstructions/models we find that the *N*-band is flipped 180° relative to the *LM*-bands. We correct all images and models in this paper for this artificial rotation.

## 5 Calibrated V$^2$ and closure phase

From the accepted exposures, we obtained 48 independent squared visibility measurements (24 in each band) and 16 (*LM*-band) + 17 (*N*-band) closure phases.

Extended Data Figures 4 and 6 plot the calibrated squared visibilities and closure phases as a function of wavelength, along with model values provided by the multi-Gaussian fits, described in §6.5, for the same *uv*-positions.

## 6 Morphological Analysis of the data

### 6.1 Introduction

Extended Data Figures 4 and 6 demonstrate that the visibilities vary strongly with *uv*-position, reaching near-zero values at position angles near 80°. This is a clear sign that we are resolving the object. Additionally, the closure phases display large non-zero values varying strongly with wavelength and with very different behaviours among the four different triplets of telescopes. This indicates that the target is strongly asymmetric.

Historically, optical/infrared long baseline interferometry has yielded too few *uv*-points to provide the direct image synthesis that is common at radio wavelengths, but recent instruments such as GRAVITY and MATISSE were designed to collect *uv*-data more efficiently. Experiments with poor or sparse *uv*-coverage are usually analysed by parameterizing rather limited models to fit the data; from *uv*-rich data direct images can be examined. Our current data set is of intermediate richness and invites experimentation into the optimum methods for extracting physical quantities. In this section we present four methods that we applied to obtain detailed morphology and reliable fluxes from the instrumental measures: MIRA[17], IRBis[16], a multi-Gaussian model, and a point-source model[41].

### 6.2 Interferometric image reconstruction

Aperture-synthesis imaging with optical/infrared arrays, such as the ESO-VLTI, is an ill-posed inverse problem. Images are reconstructed by minimizing a cost function which includes both data and regularisation terms. The data term contains the measured data (e.g., closure phases and visibilities), and the regularisation term consists of some *a priori* assumptions about the target, e.g., positivity, smoothness. In the last three decades, several image reconstruction algorithms were developed for aperture-synthesis imaging with infrared arrays. The performance of these algorithms was evaluated in several blind tests initiated by the IAU Working Group on Optical/IR Interferometry[42-45].

In this paper, image reconstruction of the central thermal dust emission of NGC 1068 were performed with two software packages, MIRA and IRBis, in order to verify the reconstruction results. The details of the packages and the application to our data are described below.

## 6.3 Image reconstruction with IRBis

The IRBis software package is part of the MATISSE data reduction pipeline. For the reconstruction process we tested the two built-in minimization engines, the three cost functions, and the six regularization functions[16].

The current version of the MATISSE image reconstruction package IRBis (mat_cal_imarec) consists of two optimization routines: a) a large-scale bound-unconstrained nonlinear optimization routine called ASA_CG[46] (engine 1) and b) the new built-in optimization routine, a limited memory algorithm for bound-constrained optimization called L-BFGS-B[47,48] (engine 2). The version of IRBis described in ref. 16 contains two different data terms in the cost function. The errors of the calibrated data used in the image reconstruction were derived from the scatter of the different data sets recorded during one observation, each lasting about 10 min. All data sets of each observation were used separately. The starting image of each reconstruction attempt was a circular Gaussian, fit to the measured visibilities of the target. Two priors were tested: a) this fitted Gaussian and b) a constant. From the many reconstructions that were calculated, the optimal was selected using the image quality measure, "qrec." The quality measure qrec consists of the reduced-$\chi^2$ values of the closure phases and visibilities and a so-called residual ratio. Because of the sparse and inhomogeneous uv-coverage of the data, it was helpful for the image reconstruction process to reduce the weight measures at uv-points in dense clusters relative to measures in low density regions of the uv-coverage. This was done with different powers of the inverse uv-density.

### 6.3.1 *LM*-band

The field-of-view (FOV) and pixel grid used for the reconstructions in *L*- and *M*-band was 340x340 mas and 256x256 pixels, respectively. Image reconstruction experiments were performed in the wavelength ranges 3.05–3.4 μm, 3.4–4.0 μm, and 4.55–4.91 μm. The radius of the applied image space mask was increasing between 70 mas and 100 mas during each reconstruction run. Because of the large pixel number, only engine 2 (L-BFGS-B) was used because engine 1 (ASA_CG) is computationally expensive for large numbers of pixels. Main Figure 1 and ED Figure 7 show the reconstructions obtained from data recorded with the UTs. The reconstruction parameters and quality are listed in Extended Data Table 1. The reconstructions were not convolved with the interferometric resolution and have spatial resolutions of $\sim\lambda/2B$ (where B is the length of the longest baseline observed), corresponding to 2.6 mas, 3.0 mas and 3.8 mas, respectively.

### 6.3.2 *N*-band

For the *N*-band reconstruction, the FOV and pixel grid used were 360x360 mas and 128x128 pixels, respectively. The image reconstructions were performed in the wavelength range 8.0—9.0 μm, 10.0—11.0 μm, and 11.5—12.5 μm. The radius of the space mask was increasing between 160 and 184 mas during each reconstruction run. ED Figure 7 shows the *N*-band reconstructions obtained from data recorded with the UTs. The reconstruction parameters and quality are listed in Extended Data Table 1. The reconstructions are not convolved with the interferometric resolution and have spatial resolutions of $\sim\lambda/2B$, corresponding to 6.7 mas, 8.3 mas and 9.5 mas, respectively.

### 6.3.3 Estimates of IRBis image artefacts

In order to quantify the reliability of the features seen in reconstructions shown in Main Figure 1, we performed an analogous reconstruction on an artificial model. The artificial model consisted of seven gaussians similar to our Multi-gaussian model for the galaxy (§6.5 below). We created visibility and closure phase data for this model with the same uv-coverage as our galaxy data, and similar noise. We performed a reconstruction with IRBis with identical regularization models and parameters. Images of the input model, the reconstruction and the residuals (reconstruction-input) are shown in Extended Data Figure 8. This process indicated that most of the artefacts visible in Main Figure 1 are caused by the limited uv-coverage, rather than by noise, and that the r.m.s. values of these artefacts was ~2% of the peak brightness. In Main Figures 1abcd we have drawn white contours at 6% of the peak. Features brighter than this almost certainly represent true sky emission.

## 6.4 Image reconstruction with MIRA

The MIRA image reconstruction package is developed and distributed by E. Thiebaut[17]. It works directly on closure phases and squared visibilities, computing $\chi^2$ for each quantity by using an approximation on the noise model that linearize these quantities in the complex plane. The minimization routine in MIRA, VMLM-B, is a variable metric method with limited memory and bound constraints on the input parameters, which is basically a variant of the BFGS algorithm.

MIRA does not introduce weighting in the *uv*-coverage as in IRBIS, nor a global parameter-search process. No weight can result in many artefacts in the reconstruction process when using a very sparse *uv*-coverage. Therefore, two methods were used to reduce and assess these artefacts in the reconstructed images.

Low Frequency Filling (LFF[49]) was used to extrapolate the visibilities at low frequencies, where the visibilities variations are deterministic[50]. This method usually strongly reduces the "ghost images" artefacts in the image reconstruction process. We experimented with several LFF sizes, and used at the end a size of 80 mas. We reconstructed images in the same central wavelengths and bandwidth as with IRBis for a direct comparison of the result.

The MIRA images shown in ED Figure 7 were reconstructed using a Total Variation regulariser, hyperparameter value of 10,000, 128 pixels, and a pixel size depending on the wavelength: 1 mas for L band, and 3 mas for N band. The start image was random, and the prior image was a Gaussian with a Full Width Half Maximum (FWHM) of 32 mas (*LM*-bands) and 96 mas *(N*-band).

The second method we experimented with was to reconstruct images at each MATISSE provided wavelength independently and computing a median image across a defined band, selected for the previous MIRA and IRBis reconstruction. This method is known to vastly reduce, but not completely remove, the typical image reconstruction artefacts that arise from a sparse *uv*-coverage. This method helped us identify the structures we consider as "real" from other structures, that clearly come from *uv*-coverage artefacts. We show a comparison of broadband MIRA reconstructions in the 3.4—4.0 µm band in ED Figure 9. The second image displays the median of a series of narrow band images within the total band, the others show images with constructed with different regularisers that attempted to fit all wavelengths in the band simultaneously, taking account of the different effective resolution at each band. The median image better matches the features found in other bands and by other reconstruction techniques.

## 6.5 Multi-Gaussian model fitting

We adapted the well-established method of Gaussian model fitting directly in the *uv*-plane to estimate fluxes. This is analogous to the method used in previous MIDI analysis of NGC 1068[51].

There are several differences in this work: we fit various components based on the shape of the main sources found through image reconstruction; we use squared visibilities and closure phases; and that we use a more sophisticated Markov Chain Monte Carlo (MCMC) Bayesian methodology to estimate parameters and their uncertainties[52]. We fit independent models to a selection of wavelengths across the entire wavelength range provided by MATISSE. Unconstrained fitting with many parameters leads to low residuals, but also to poorly conditioned and degenerate solutions. To avoid this, we imposed several constraints. First, the positions and shapes of the components had to approximate the images generated by the reconstructions. Secondly, we enforced local continuity in wavelength by making simultaneous fits at 2 (in *L*-band) or 3 (in *M*-band) different wavelengths, 0.05 µm apart along a large range of wavelengths.

We determined the minimum number of Gaussians in each band which resulted in both closure phase residuals < 10° and squared visibility residuals less than 0.1 and still visually resemble the morphology revealed in the image reconstruction. For the *LM* bands we needed 7 Gaussians to recover the structures; for the *N*-band we only needed 3 Gaussians. Each Gaussian has six free parameters: its *l,m* position, relative flux, FWHM major- and minor-axes, and the position angle of its major axis. The MATISSE visibilities and closure phases are not sensitive to the absolute position of the source complex on the sky or to the sum of the fluxes of the components. To remove these non-measured degrees of freedom from the model, we fix one Gaussian component to the centre of the image, while the positions of the centre of the other components are specified by their distances and position angles relative to the first. We also define the central brightness of each component relative to the fixed Gaussian.

We calculate the squared visibilities and closure phases from the best-fit model intensity distribution and estimate uncertainties for these quantities from the parameter posterior probability distributions. These are shown in Extended Data Figures 4 and 6. We use a similar process to estimate the uncertainties of the measured Spectral Energy Distributions (SEDs) discussed in §8. The list and definitions of fitted parameters are in Extended Data Table 1.

In Extended Data Figure 9 we show the best fitting models at all modelled wavelengths. In Extended Data Figure 7 we show the best-fitting models at four chosen wavelengths for comparison to other methods, 3.6, 4.77, 8.50, and 11.5 μm. We show the best fitting parameters for the Gaussian model at three wavelengths, one in each band, in Extended Data Table 1. In the *L*- and *M*-bands we can identify two main structures separated primarily along North-South axis; we refer to them as **NC** (Northern Complex) and **SE** (Southern Extension). In the *N*-band, at 12 μm, the distribution of the intensities given by our 3 Gaussians look similar to the two central sources of Lopez-Gonzaga's Model 1[51]. We compare the model squared visibilities and closure phases to the data in Extended Data Figures 4 and 6. The average RMS of the residuals are 8.2 ,7.6, and 8.3° in closure phase, and of 0.01, 0.01 and 0.01, in Vis$^2$, in *L*-, *M*-, and *N*-bands, respectively.

In the Gaussian modelling, we did not require a background component due to the already extended nature of the Gaussians. As a check, we performed a similar fit with an additional extended component, but in all cases the resulting residuals were larger. This indicates that the majority of the flux comes from the Gaussians in our models at the different wavelengths. We therefore distribute the single-dish flux among all the components fitted at each wavelength. When determining the photometry, the value and uncertainties are calculated by bootstrapping, with replacement, the last 4000 samples of the associated MCMC chain.

## 6.6 Point source model fitting

We further model the interferometric data using a point source modelling method analogous to clean-like image reconstruction. The point source model allows us to create an independent model image of the data, to be compared to the image reconstruction, while also providing photometry at the same narrow wavelength bins as the Gaussian modelling.

### 6.6.1 The point source model

The base model fitting method is described in previous work[41]. We use the method which fits an image consisting of $N_p$ point sources to the squared visibility and closure phase data. Unlike the described model, we also include an extended background component and do not include the scale factor (which not relevant for such a bright object). We chose the extended background source to be an elongated Gaussian, unfixed in position and amplitude, with a fixed major axis FWHM of 70 mas, minor axis FWHM of 50 mas, and major axis PA of 0°. The background Gaussian's fixed parameters were determined from the *N*-band Gaussian models. Each image is 256x256 pixels, we use 200 walkers and 1500 iterations.

### 6.6.2 Verifying the image reconstruction

When determining geometry independently of the image reconstruction with the point source model, we used the random starting position option for the walkers. We separated the observed data into 0.1 μm width bins centred at 3.55 μm, 3.65 μm, 3.8 μm, 3.9 μm, 4.65 μm, and 4.8 μm. The data in these bins is not averaged during fitting but instead used to extend coverage of the *uv*-plane, i.e., assuming the different wavelengths see the same structure at a different angular resolution.

Initially, we fit the 4.8 μm bin. The first of the two fits performed when using a random starting position was performed for the $N_p$ range of 3 to 16 sources. It was found that the preferred value of $N_p$ for the chosen is bin width is 9. We used the result as the starting point of the second fit for the 4.8 μm bin as well as the other wavelength bins.

### 6.6.3 Point source photometry

The $N_p = 9$ point-source model is a good representation of the observables and sufficient to compare geometry but the discrete nature of the model introduces uncertainty to any fixed aperture photometry. To solve this, we overfitted the data. Each of the nine point-source starting positions from the geometric modelling was tripled and randomly shifted in position by a normal distribution of 0.2 mas, giving 27 point sources. An $N_p = 27$ fit for each wavelength bin was performed with this starting position. The resulting models are visually similar to the geometric models and a good description of the data. The photometry and its uncertainties at each wavelength are calculated by performing aperture photometry on a bootstrapped, with replacement, selection of the 100000 samples provided by the last 500 iterations of the 200 walkers.

## 6.7 Comparison of reconstruction methods

### 6.7.1 Morphology

We find excellent agreement in source morphology using the various modelling and reconstruction techniques. The results of the four imaging techniques at different wavelengths are shown in Extended Data Figure 7. The *L*-band images have the most substructure as a result of the smaller angular resolution, so we focus our quantitative comparison on this band. In this band, regardless of approach, we find two main features: an incomplete ellipse with a major axis PA of ~ -45° and a secondary southern source. In the image reconstructions we take a conservative uncertainty on the size measurements equal to 1 pixel (1.325 mas in IRBis; 1 mas in MIRA). The measured outer extent of the broken ellipse in each method are (minor x major): 8.1x13.8 mas with PA=-45° using IRBIS; 7.5x13.2 mas with PA=-45° using MIRA; 7.6x12.2 mas with PA=-40° using Gaussian fits; and 6.5x13.6 mas with PA=-45° using point-source reconstruction. We also measure the distance between the centre of the broken ring and the centre of the secondary source: 15.6 mas with IRBIS; 17.13 mas using MIRA; 15.3 mas using Gaussian modelling; and 17.2 mas using point-source reconstruction. We note that SE in the Gaussian modelling is much larger than for the other reconstruction methods. We hypothesize that this is a result of the simplicity of a Gaussian component, and the true southern source may have multiple components as seen in the MIRA, IRBIS, and point-source reconstructions. Nonetheless, within uncertainties the dimensions of the broken ring and the separation of the sources agree remarkably well, especially considering our sparse *uv*-coverage. We see a similar level of consistency within the *M*-band and between the two bands. More qualitatively, we find less agreement in the *N*-band, especially at longer wavelengths where MIDI revealed significant extended flux to the North[51]. With >30 m baselines and the smaller field-of-view of the UTs relative to the MIDI AT observations, we expect to resolve-out this extended component, but we cannot rule out that it contributes to the MATISSE *N*-band flux. In particular, this extended component can be broken up into "noise"-peaks in image reconstruction. With this in mind, we find a consistent PA=-45° bright central structure in each *N*-band reconstruction. It is accompanied in each case by extended flux: to the north in both MIRA and Gaussian modelling; and roughly evenly north-south in the IRBIS reconstruction and point-source reconstruction. We can conclude only that the central bright source is consistent and that it must be accompanied by extended flux. At 8.5 µm where this extended background is less prominent, we find excellent agreement between the techniques: a central ~15 mas/1 pc component with PA=-45° flanked to the north and south by fainter extended flux. In the IRBIS images, we see hints of the broken ring continuing from the *LM*-band. As we see consistent morphology across both methods and wavelengths, we report the first images of this inner region of NGC 1068 with high confidence.

### 6.7.2 Photometry

Here we will compare the photometry from both models and the image reconstructions. We compare the photometry for the apertures: E1 to E5, DE3, SE, DE1 with NC contributions removed, and DE2-SE (see Main Figure 2). The results are shown in ED Figure 2. The photometry shown is the relative flux, under the assumption that all flux in the image sums to one. Because the apertures do not cover the whole image, all apertures for one image may not sum to one.

In the NC (E1-E5), we find good agreement in all ellipses with all methods. There is a slight discrepancy in the IRBis flux in E3, primarily in *M*-band. Overall, we find that the photometries from the independent methods are similar in spectral shape and relative flux. The similar spectral shape will lead to similar intrinsic temperatures, absorption signatures, and chemistry. This suggests that the photometry used for the associated SEDs is reliable and mostly insensitive to methodology.

For the SE, we find good agreement between the images and point source model, but the Gaussian model shows much larger fluxes in *L*- and *M*-band. We tested whether this was a background effect by introducing a background component to the Gaussian model that matched the one included in the point source model. We found that this results in a poorer fit and no significant change in the *LM* fluxes. When we use the background-included model fluxes for the SED fitting (assuming no carbon), we find similar temperatures. Furthermore, the inclusion of the background did not affect the extracted photometry of the NC. Finally, we also fit an SED, assuming no carbon, created from the SE photometry of the point source model. We find a hot component temperature of 808—846 K instead of 882—974 K which is cooler but not in conflict with the results. We hypothesize that the difference between methods for the SE is due to a limitation of Gaussian modelling, i.e., the SE is complex and not well described by a Gaussian. For the Diffuse Emission (DE) apertures, we find relatively large scatter. While more scattered, the DE apertures have similar overall spectral shapes between methods.

In conclusion, we find that the aperture photometry is suitably reliable for our scientific undertaking and structurally consistent between the methods presented in this work. While extracting the fluxes from any method should produce suitable SEDs for the temperature modelling, we choose to use the Gaussian modelling. Using a model instead of a reconstruction allows us to easily extrapolate through the model and data uncertainties using the derived parameter distributions from MCMC modelling as well allowing narrower wavelength bands for more detailed SEDs. The Gaussian model specifically is the most comparable to previous studies of NGC 1068 with MIDI[12-14]. Furthermore, it provides fine control over the brightness distribution which is useful when determining the importance of revealed structures.

# 7 Radio data

The nucleus of NGC 1068 was observed using the High Sensitivity Array (HSA), which includes the Very Long Baseline Array (VLBA), the phased Very Large Array (VLA), and the Green Bank Telescope (GBT) (Gallimore, J. F. and Impellizzeri, C.M.V, in preparation). The observations took place on 8—9 February 2020 and 21—22 March 2020. The receivers were tuned to the 22.235080 GHz transition of water vapor, offset to the redshift of NGC 1068. The details of the observations, and particularly the analysis of the detected $H_2O$ masers will be published separately. Data reduction, including bandpass calibration, fringe-finding, and aperture synthesis imaging were performed using standard techniques in AIPS[53-56]. The final calibration of interferometric phase was based on the brightest (few hundred mJy) maser sources of NGC 1068, and phase solutions were transferred to the nearby astrometric calibrator J0239-02, located 2.°7 from NGC 1068. By evaluating the offset sky positions on images of the astrometric calibrator, the authors were able to determine the absolute positions of the NGC 1068 maser sources to an accuracy of about 0.3 mas.

The final products are a data cube with axes of sky coordinates and radial velocity and a single-channel image of 22 GHz radio continuum. Continuum was detected in line-free channels only on the shortest interferometric baselines. To produce the continuum image, the authors averaged the line-free channels at the ends of the observed bandwidth and applied a Gaussian taper with 50% weight at 30 M$\lambda$ during Fourier inversion. The resulting continuum image was then deconvolved using the multiscale CLEAN algorithm[57]. and restored with a Gaussian beam of 4.3x3.3 mas, at PA -21°. This image is displayed as contours in Main Figure 4b. The peak brightness in this image is 3.0 mJy/beam. The total flux density of the recovered 22 GHz continuum is $S_\nu$=13.8±0.3 mJy. The centroid position of the resolved continuum is $\alpha_{J2000}$= $02^h42^m40^s.70901$, $\delta_{J2000}$ = 00°00"47".9448.

This bright central radio emission is almost certainly free-free radiation from gas hotter than $10^6$ K because of its flat spectrum between 5 and 22 GHz, its brightness temperature of >2 x $10^6$ K at 5 GHz and its inverted spectrum below this frequency[24]. The nature of the diffuse 256 GHz emission[22] seen in Main Figure 4(a) is less clear; it may be a combination of free-free emission and the long-wavelength tail of the infrared dust emission. The total 256 GHz flux is 12.7±0.1 mJy, composed of an unresolved point of 6.6 mJy and a diffuse component of 6.1 mJy[22]. The peak brightness in Fig. 4(a) is 0.85 mJy/beam. The estimated total free-free flux, extrapolated from lower frequencies is 13±1 mJy similar to the measured total flux[58]. The blackbody fits of the diffuse *N*-band infrared emission (§8 below) can be extrapolated to 256 GHz (1.2 mm) and lead to an estimated flux of 30 mJy, which exceeds the measured flux, but this assumes that the emitting dust is optically thick even at millimetre wavelengths. This requires a large number of very large, i.e. millimetre size, dust grains, which seems unlikely. Galactic dust particles have emission efficiencies that scale at large wavelengths ~$\lambda^{-1}$ (c.f. [59]) Thus if the 12 $\mu$m emission optical depth is not >>1, the expected 1.2 mm flux would be about 100 times smaller than the blackbody extrapolation. From these considerations we believe that the diffuse emission too arises most likely from hot gas rather than warm dust.

### 7.1 Radio-Infrared relative positions

We determined the position of the VLBA 22 GHz peak relative to the 3.7 $\mu$m peak by a three step process. First, we determined the displacement of the ALMA 256 GHz image (Main Fig. 4a) with respect to the MATISSE 12 $\mu$m image by cross-correlating the images. The formal error in this process, essentially the width of the cross-correlation function divided by its Signal/Noise ratio, was 1.6 mas. Secondly, we assume the MATISSE 12 $\mu$m peak coincides with the MATISSE 3.7 $\mu$m peak. The relative positions can be determined to an accuracy of <3 mas. Lastly, we can align the VLBA 22 GHz image with the ALMA 256 GHz map to <1 mas, based on their absolute astrometry. The three steps yield Fig. 4b, where the relative positions are accurate to ~3 mas.

## 8 Spectral Energy Distribution Modelling

### 8.1 SED Estimates

In §6.5, we produced Gaussian models that well reproduced the observed squared visibilities and closure phases. We multiplied the relative fluxes of each aperture (see ED Figure 2) in each band by the photometric flux densities determined in Methods §4 (ED Figure 5) for the same wavelength range. This results in the actual flux densities of the components in these bands. For the uncertainties, we propagated the derived error from the Gaussian model fluxes and the observed photometry. To determine the distribution of dust temperature, we defined 9 main areas of interest (shown in Main Fig. 2), 5 small ellipses over the structure of NC, 1 medium ellipse over SE, and 3 large ellipses to encompass the extended emission that predominates at larger wavelengths. The small areas were defined to cover the features seen in the L and M band multi-Gaussian models, with as many ellipses as possible but larger than the resolution limit in the N band, so we could integrate their fluxes over the three bands. The large ellipses (DE1 to DE3) were defined to cover the remaining areas with the extended and/or diffuse emission in the N band multi-Gaussian models, based on the narrow band models, so to recover any spectral changes along the N band. Since the N band multi-Gaussian models only used three components we matched that number in ellipses to not to cause an over-sampling. Because of this, the latter do not perfectly gather all the flux or emitting areas at every wavelength, and also because the structures change substantially along the N band due to the Si feature. We then summed the contribution from each Gaussian component in each of the defined areas to obtain Spectral Energy Distributions (SEDs). The plotted points with error bars in Extended Data Figure 3 and Main Figure 3 represent the estimated SEDs in the nine selected apertures.

## 8.2 SED Fitting

For each aperture we fit the formula given in the SED section of the Main paper with one blackbody where this was sufficient, and otherwise with two blackbodies. For each set of measured fluxes, we performed a brute force evaluation of $\chi^2$ over the range of parameters in the formula. We then determined the minimum value $\chi^2_{min}$, and the parameter envelope where $\chi^2 - \chi^2_{min} < \sqrt{2n_p}$

$n_p = 6$ is the number of fitted parameters. For a non-linear, multi-parameter model, this envelope corresponds in confidence to the 1-σ level for a single gaussian variable. The shaded areas in ED Figure 3 and Main Figure 3 indicate the union of all SEDs of the models that met this criterium.

These SEDs rely on the cross-identification that we make of the Gaussian models, or images, from wavelength to wavelength, which is based on overlapping the brightest points at all wavelengths. Further details are given in the main article.

### 8.2.1 Opacity laws and dust mineralogy

The thermal modelling process requires the definition of the foreground opacity law, $\kappa(\lambda)$. The standard Interstellar Medium (ISM) extinction profile with a Mathis-Rumpl-Nordsieck (MRN) size distribution[59], which resembles the extinction measured toward the Galactic Center[20] resulted in bad fits to the silicate absorption feature and the fluxes at 8 μm. (c.f. Main Figure 3). We then tried fitting κ with a series of theoretical curves[21] modelling amorphous silicate grains of various sizes with various ratios of pyroxene to olivine and of Iron to Magnesium. Dust containing significant amounts of pyroxene shows excessive absorption at 8 μm. The same is true for particle sizes > ~5 μm and iron-poor silicate mixtures. We conclude that the absorbing material in front of the hot emitting regions contains predominantly Fe/Mg olivine particles of small to moderate size (0.1—1.5 μm). Due to the 3.4 μm carbonaceous features present in the L band, we additionally tested varying amounts of amorphous carbon particles of the same size.

### 8.2.2 Two-temperature model results

We show in Main Figure 3 and Extended Data Figure 3 the results of the 2-component blackbody fits to the SEDs of the ellipses, for models using no carbon and models including 20% of carbon by weight. Larger carbon contribution produced significantly poorer fits. As more carbon is added, the temperatures of the hot component rise. We list in Extended Data Table 1 the ranges of the values of the parameters for the models with fits inside 1σ at each wavelength. For the fits we set a maximum $T_{hot}$ of 1500 K, a minimum $\eta_{hot}$ of $10^{-5}$, (below that the contribution of the flux is irrelevant) and a minimum $N_{ext}$ of 1μg cm$^{-2}$ (lower values become degenerate with the combination of high temperatures of ~1000 K). There is some cross-talk between the parameters $N_{ext}$ and $T_{hot}$ from around 900 K, i.e., if we use a higher temperature for the hot component then we have to add a larger amount of extinction.

## 8.3 Dust Composition

The composition, crystallinity and size distribution of the dust grains contain clues to their origin and thermal history. These in turn contain clues about the accretion processes for dust and gas. The *N-band* silicate feature, discussed in §8.2.1 is striking in its weak opacity near 8 μm, in contrast to "standard interstellar" curves or the opacity toward the Galactic Centre. In our Galaxy, low 8 μm opacity characterizes the diffuse interstellar medium (ISM), rather than dust in dense molecular clouds[20] where grains are larger and have icy mantels that add opacity at this wavelength. Our opacity curve in fact most closely resembles Galactic extinction curves measured in low density regions far from the centre[60]. The comparison to theoretical curves indicates a higher olivine/pyroxene ratio than in the general Galactic diffuse medium, but the possible relation of this ratio to the ISM density or distance to the Galactic Centre is not well studied. Some authors[61] have attributed the 8 μm peak to SiC grains, and the 3.4 μm C-H feature may hint at some Carbon rich absorbing grains, but the general good match to olivine rich silicates, and the crystalline features seen in higher-resolution spectra (§8.4, Extended Figure 5) favour the silicate interpretation.

## 8.4 Thermally reprocessed silicates

During the 24 Sept. 2018 commissioning run, we took a MATISSE snapshot using the N-band HIGH resolution grating (R=300). At this resolution some of the baselines showed a flattened, double-peaked, silicate absorption feature (Extended Data Figure 5). This spectral form indicates that the grains are crystalline rather than amorphous. Analogous to the grains near Young Stellar Objects (YSOs) they have probably been melted and recrystallised in a hot environment[55]. This feature can also be seen in previous published high-resolution MIDI data[7].

## Further References

32. Leinert, C. *et al.* MIDI – the 10 μm instrument on the VLTI. *Astrophys. Space Sci.* **286**, 73–83 (2003).
33. Cruzalèbes, P. *et al.* A catalogue of stellar diameters and fluxes for mid-infrared interferometry. *Mon. Not. R. Astron. Soc.* **490**, 3158–3176 (2019).
34. Burtscher, L., Tristram, K. R. W., Jaffe, W. J. & Meisenheimer, K. Observing faint targets with MIDI at the VLTI: the MIDI AGN large programme experience. in *Optical and Infrared Interferometry III* vol. **8445** 494–506 (SPIE, 2012).
35. Jaffe, W. J. Coherent fringe tracking and visibility estimation for MIDI. in *New Frontiers in Stellar Interferometry* vol. **5491** 715–724 (SPIE, 2004).
36. Millour, F. *et al.* Data reduction for the AMBER instrument. in *New Frontiers in Stellar Interferometry* vol. **5491** 1222–1230 (SPIE, 2004).
37. Cohen, M. *et al.* Spectral Irradiance Calibration in the Infrared. X. A Self-Consistent Radiometric All-Sky Network of Absolutely Calibrated Stellar Spectra. *Astron. J.* **117**, 1864, 1864 (1999).
38. Petrov, R. G. *et al.* Commissioning MATISSE: operation and performances. in *Optical and Infrared Interferometry and Imaging VII* vol. **11446** 124–142 (SPIE, 2020).
39. Bourges, L. *et al.* VizieR Online Data Catalog: JMMC Stellar Diameters Catalogue - JSDC. Version 2 (Bourges+, 2017). *VizieR Online Data Catalog* **2346**, (2017).
40. Meilland, A. *et al.* The binary Be star δ Scorpii at high spectral and spatial resolution. I. Disk geometry and kinematics before the 2011 periastron. *Astron. Astrophys.* **532**, A80 (2011).
41. Leftley, J. H. *et al.* Resolving the Hot Dust Disk of ESO323-G77. *Astrophys. J.* **912**, 96 (2021).
42. Lawson, P. R. *et al.* An interferometry imaging beauty contest. in *New Frontiers in Stellar Interferometry* vol. **5491** 886–899 (SPIE, 2004).
43. Cotton, W. *et al.* 2008 imaging beauty contest. in *Optical and Infrared Interferometry* vol. **7013** 531–544 (SPIE, 2008).

44. Baron, F. *et al.* The 2012 interferometric imaging beauty contest. in *Optical and Infrared Interferometry III* vol. **8445** 470–483 (SPIE, 2012).
45. Sanchez-Bermudez, J. *et al.* The 2016 interferometric imaging beauty contest. in *Optical and Infrared Interferometry and Imaging V* vol. **9907** 372–389 (SPIE, 2016).
46. Hager, W. W. & Park, S. Global convergence of SSM for minimizing a quadratic over a sphere. *Math. Comput.* **74**, 1413–1423 (2005).
47. Byrd, R. H., Lu, P., Nocedal, J. & Zhu, C. A limited memory algorithm for bound constrained optimization. *SIAM J. Sci. Comput.* **16**, 1190–1208 (1995).
48. Zhu, C., Byrd, R. H., Lu, P. & Nocedal, J. Algorithm 778: L-BFGS-B: Fortran subroutines for large-scale bound-constrained optimization. *ACM Trans. Math. Softw.* **23**, 550–560 (1997).
49. Millour, F. *et al.* AMBER on the VLTI: data processing and calibration issues. in *ESO calibration workshop 2007* (ed. ESO) (2007).
50. Lachaume, R. On marginally resolved objects in optical interferometry. *Astron. Astrophys.* **400**, 795–803 (2003).
51. López-Gonzaga, N., Jaffe, W., Burtscher, L., Tristram, K. R. W. & Meisenheimer, K. Revealing the large nuclear dust structures in NGC 1068 with MIDI/VLTI. *Astron. Astrophys.* **565**, A71 (2014).
52. Foreman-Mackey, D., Hogg, D. W., Lang, D. & Goodman, J. emcee: The MCMC Hammer. *Publ. Astron. Soc. Pacif.* **125**, 306 (2013).
53. Wells, D. C. NRAO'S Astronomical Image Processing System (AIPS). in *Data Analysis in Astronomy* (eds. Gesù, V. D., Scarsi, L., Crane, P., Friedman, J. H. & Levialdi, S.) 195–209 (Springer US, 1985). doi:10.1007/978-1-4615-9433-8_18.
54. Wells, D. C. NRAO's Astronomical Image Processing System (AIPS). 195 (1985).
55. Greisen, E. W. The Astronomical Image Processing System. in *Acquisition, Processing and Archiving of Astronomical Images* (eds. Longo, G. & Sedmak, G.) 125–142 (Osservatorio Astronomico di Capodimonte & FORMEZ, 1988).
56. Greisen, E. W. AIPS, the VLA, and the VLBA. in *Information Handling in Astronomy - Historical Vistas* (ed. Heck, A.) 109–125 (Springer Netherlands, 2003). doi:10.1007/0-306-48080-8_7.
57. Cornwell, T. J. Multiscale CLEAN Deconvolution of Radio Synthesis Images. *IEEE Journal of Selected Topics in Signal Processing* **2**, 793–801 (2008).
58. Cotton, W. D., Jaffe, W., Perrin, G. & Woillez, J. Observations of the inner jet in NGC 1068 at 43 GHz. *Astron. Astrophys.* **477**, 517–520 (2008).
59. Mathis, J. S., Rumpl, W. & Nordsieck, K. H. The size distribution of interstellar grains. *Astrophys. J.* **217**, 425–433 (1977).
60. Zasowski, G. *et al.* Lifting the Dusty Veil with Near and Mid-IR Photometry II: A large-scale study of the Galactic Infrared Extinction Law. *Astrophys J.* **707**, 510-523 (2009)
61. Köhler, M. & Li, A. On the anomalous silicate absorption feature of the prototypical Seyfert 2 galaxy NGC1068. *Mon. Not. R. Astron. Soc.* **406**, L6–L10 (2010).
62. Van Boekel, R et al, The building blocks of planets within the `terrestrial' region of protoplanetary disks. *Nature* **432**, 479-482 (2004)
63. Isbell, J. W. *et al.* Subarcsecond Mid-infrared View of Local Active Galactic Nuclei. IV. The L- and M-band Imaging Atlas. *Astrophys. J.* **910**, 104 (2021).
64. Prieto, M. A. *et al.* The spectral energy distribution of the central parsecs of the nearest AGN. *Mon. Not. R. Astron. Soc.* **402**, 724–744 (2010).

## Acknowledgements

The authors thank ESO and particularly the Cerro Paranal staff for their support in obtaining these observations. The data presented here was taken as part of ESO projects 60.A-9257(commissioning) and 0104.B-0322(A) (MATISSE Guaranteed Time Observations of AGNs).

We thank the GRAVITY AGN team for useful scientific discussions and early access to digital versions of their data.

MATISSE was defined, funded, and built in close collaboration with ESO, by a consortium composed of French (INSU-CNRS in Paris and OCA in Nice), German (MPIA, MPIfR and University of Kiel), Dutch (NOVA and University of Leiden), and Austrian (University of Vienna) institutes. The Conseil Départemental des Alpes-Maritimes in France, the Konkoly Observatory and Cologne University have also provided resources for the manufacture of the instrument. A thought goes to our two deceased OCA colleagues, Olivier Chesneau and Michel Dugué, with us at the origin of the MATISSE project, and with whom we shared many beautiful moments.

VGR was partially supported by the Netherlands Organisation for Scientific Research (NWO). JHL acknowledges the support of the French government through the UCA JEDI investment in the Future project managed by the National Research Agency (ANR) under the reference number ANR-15-IDEX-01.

Contributions:

VGR, JWI, WJ, RP, K-HH, FM, JL, AMe: Observing, data reduction, calibration, modelling, interpretation

BL, SL, FA, SR-D, PC, PB, FB, TH, GW, PA, UB, UG, MH, ML, AMa, DS, PS, JW, GZ, PB: MATISSE instrument design, fabrication, and commissioning, calibration

LB, GW, RvB, PS, J-CA, MH, J-UP: Scientific Planning

WCD, JFG, CMVI, KT, LB, CP: Observing

CD, JD, VH, JH, LK, EK, LL, EP, AS, JV, SW, LBFW, GY, WCD: Interpretation

## Data Availability:
The raw MATISSE data used in this article are available to qualified researchers at:

http://archive.eso.org/eso/eso_archive_main.html

Reduced data is available at: https://github.com/VioletaGamez/NGC1068_MATISSE

## Code Availability
The python code for the emcee sampler is available via:

https://emcee.readthedocs.io

The python code to fit Multi-Gaussian models, and Spectral Energy Distributions is available at:

Thermal imaging of dust hiding the black hole in NGC 1068 (v1.0). Zenodo.
https://doi.org/10.5281/zenodo.5599363

The MiRA image reconstruction code is available at:

https://github.com/emmt/MiRA

<>
The ESO MATISSE pipeline, including IRBis is available from https://www.eso.org/sci/software/pipelines/matisse/matisse-pipe-recipes.html

The authors have no competing interests to report.

Correspondence and requests for materials should be addressed to V. Gámez Rosas (gamez@strw.leidenuniv.nl)

Reprints and permissions information is available at www.nature.com/reprints

# Extended Data

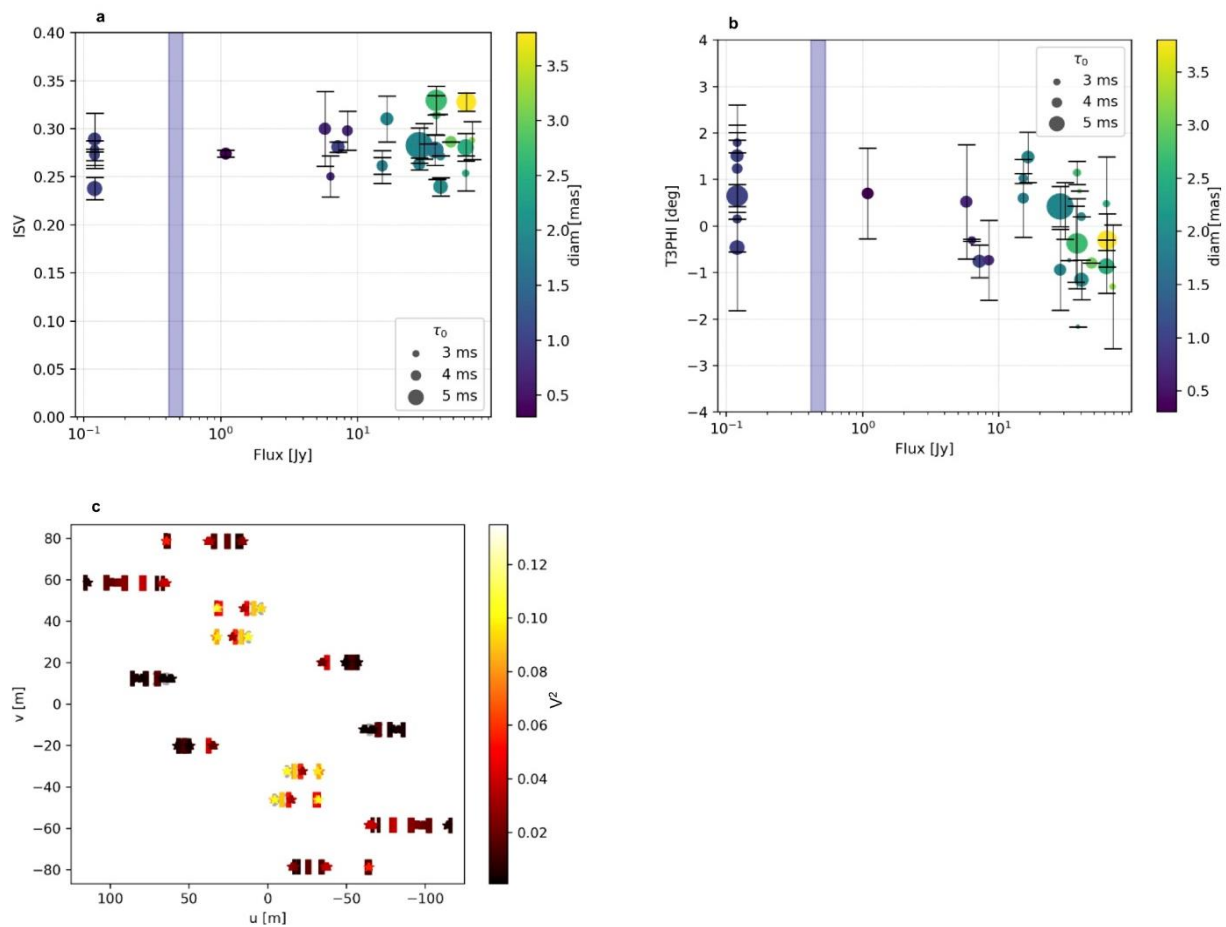

***Extended Data Figure 1. MATISSE Faint Calibrator data and uv-coverage**. **a**: Instrumental Visibility and **b**: non-calibrated closure phases of calibrators observed during the months of September 2018, May 2019 and June 2019, at 3.4 µm. Data points are colour-coded by their diameters, and the size of the circles correspond to the average coherence times, of the observations. The vertical blue strip covers the approximate correlated flux of NGC 1068 at the same wavelength.*

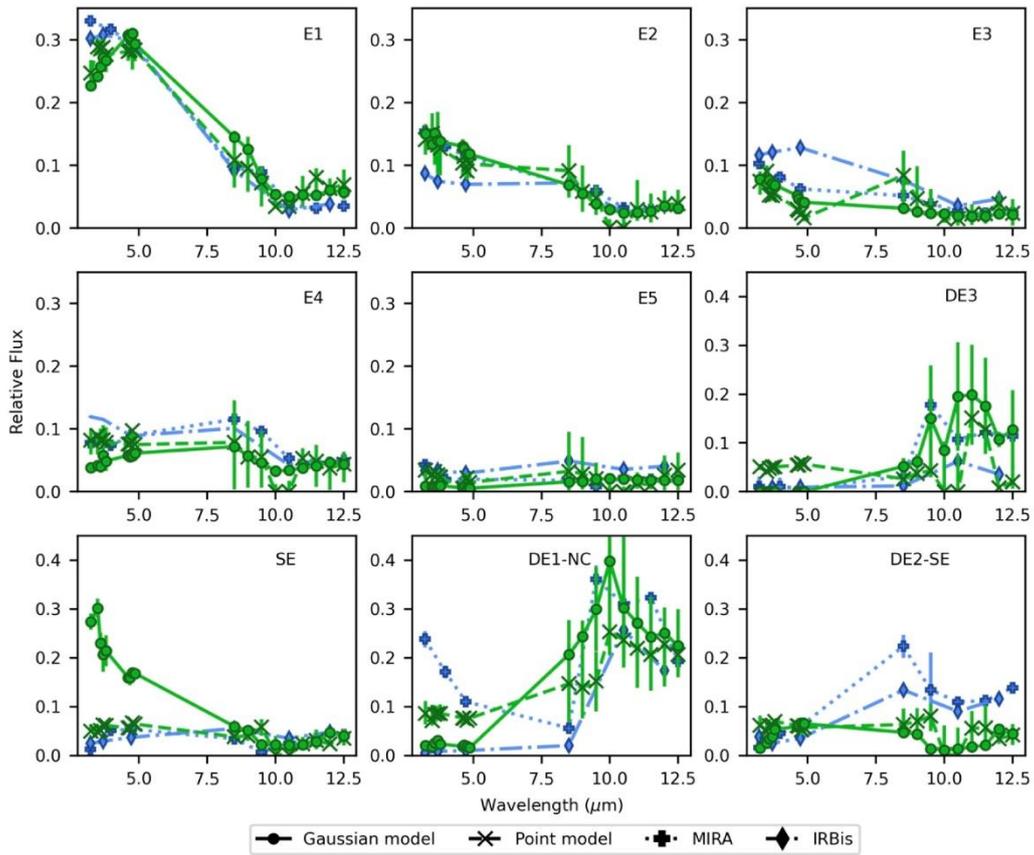

**Extended Data Figure 2: Comparison of photometry of the models and image reconstructions.** *The photometry between methods generally agrees in spectral shape between different methods. The most notable exceptions are SE and E3 which still produce similar temperatures from SED modelling between methods.*

**Extended Data Figure 3: SED black body model fits to MATISSE aperture photometry.** *The figures are labelled with the aperture names defined in Main Figure 2. The shaded areas show all models falling inside 1 sigma of the photometry, considering both pure amorphous olivine (magenta) and a mix of olivine and 20% amorphous carbon by weight (cyan). The plots for apertures E1 and SE are in the Main article.*

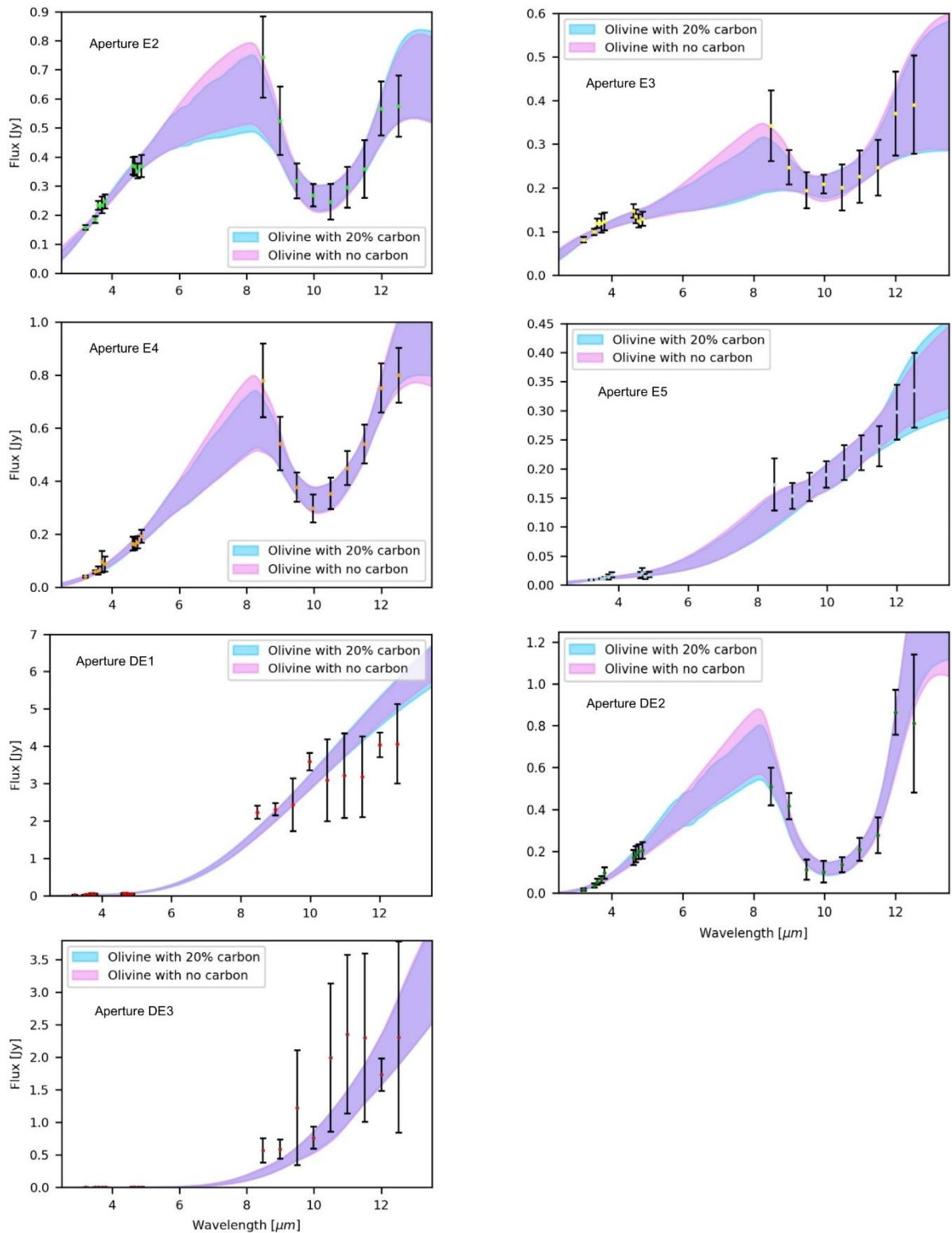

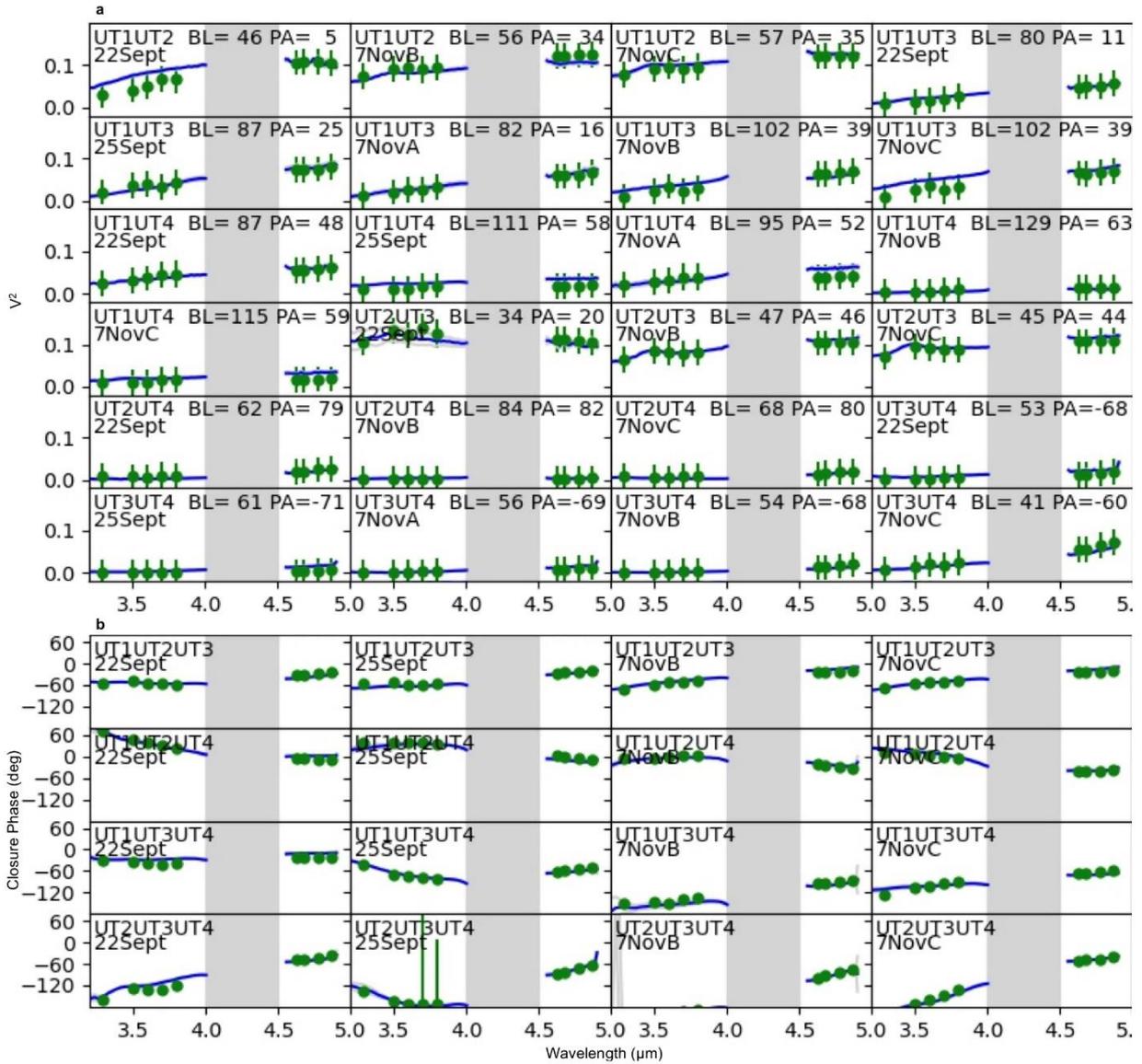

Extended Data Figure 4. NGC 1068 LM-band data compared to best multi-gaussian model.

a: Squared visibilities for NGC 1068: The blue lines show observed values, averaged over sub-exposures; the thin grey lines show individual sub-exposures in order to illustrate the measurement uncertainties, but are often hidden behind the blue lines. The green points with error bars show values predicted by the multi-Gaussian models from Methods §6.5. The error bars represent the r.m.s. sum of the measurement errors and the uncertainties of the model parameters. The distance between models and observations shows that a limited number of Gaussians cannot exactly represent the true sky or that we do not have a sufficient uv coverage and/or resolution. The grey bands mark the atmospheric non-transmission band. The labels indicate the telescope pairs for each baseline, the baseline length(m) and position angle (degrees), and the specific exposure label from the observation log described in the data access section of Main. b: closure phases (degrees) using the same colour code as above. The labels indicate the telescope triplets and the specific exposure label from the observing log.

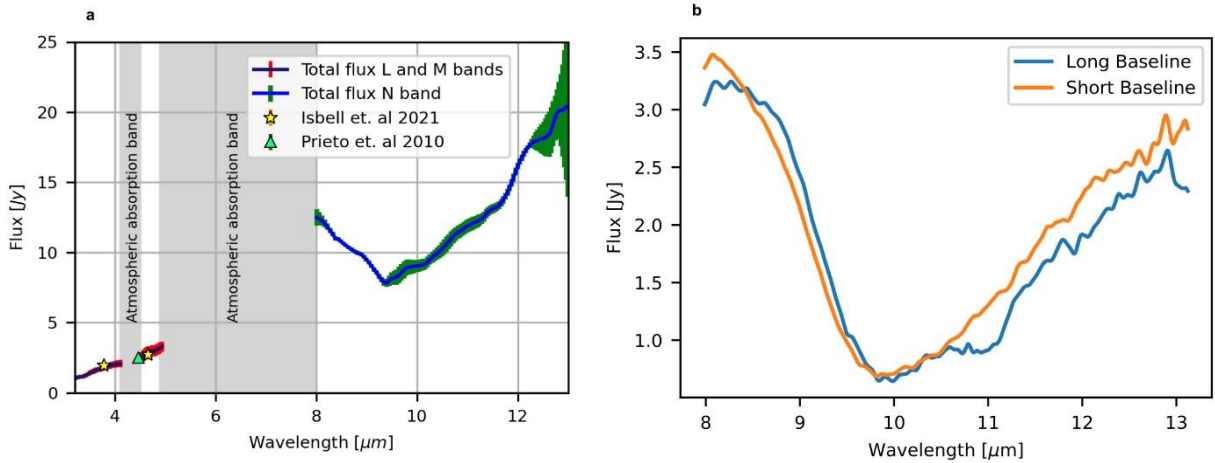

***Extended Data Figure 5: NGC 1068 spectra: a:*** *Average single telescope spectrum of NGC 1068 in LM-bands (black solid line) and N-band (blue solid line). The error bars represent uncertainties estimated from the differences between different dates and calibrators. The yellow stars refer to VLT/ISAAC L' -and M-band single-dish flux estimates from reference [63], while the green triangle corresponds to a VLT/NACO M-band flux from [64].* ***b:*** *The silicate absorption feature observed on two baselines at high spectral resolution (R~300) during a single MATISSE commissioning snapshot. The 85 m baseline shows the broader, double-peaked profile characteristic of crystalline, reprocessed grains[62]. The difference between the curves shows that the crystallinity varies over the source.*

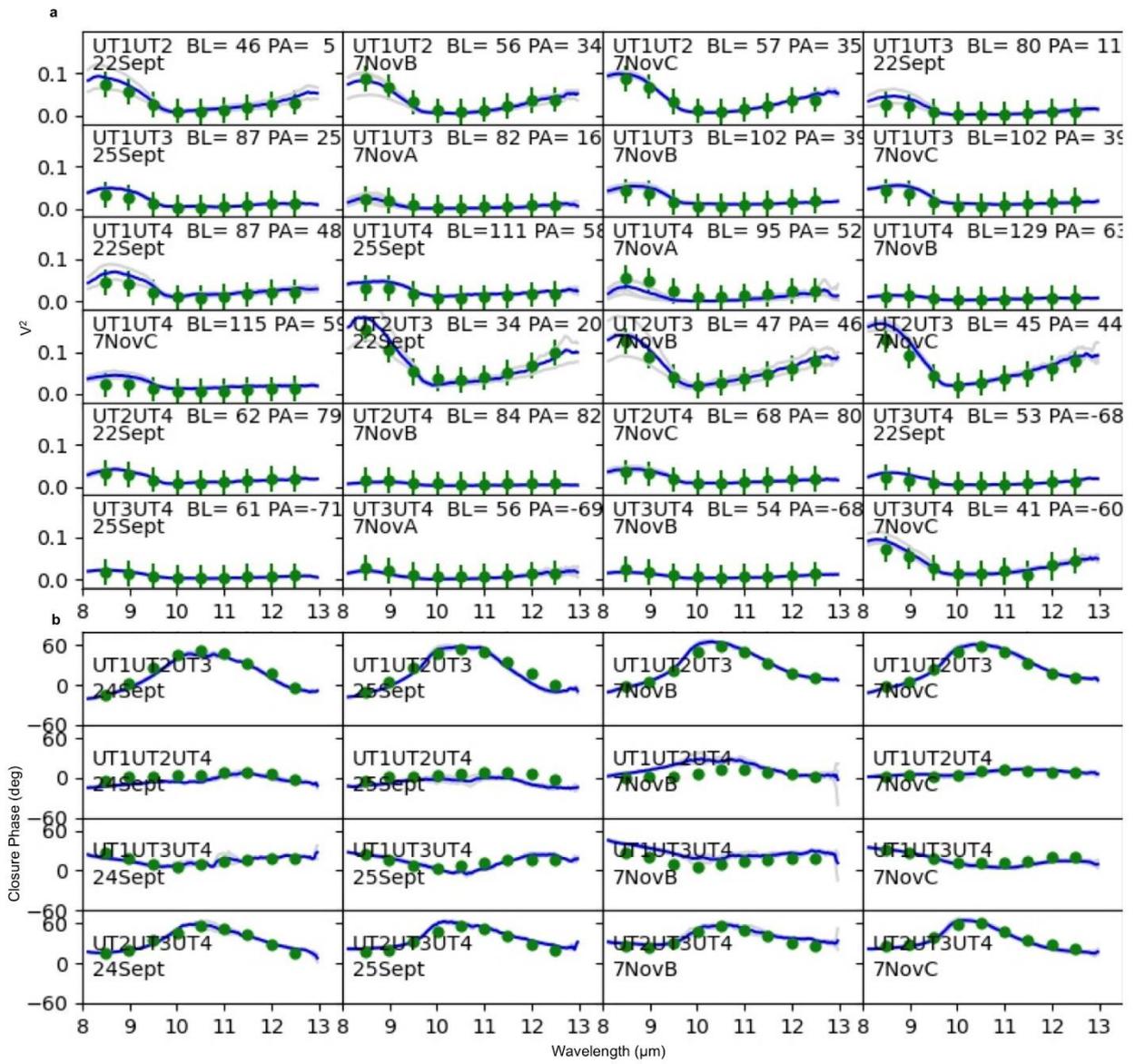

**Extended Data Figure 6: MATISSE N-band squared visibilities and closure phases.** *The quantities plotted, and the symbols used are the same as ED Figure 4 for the N band.*

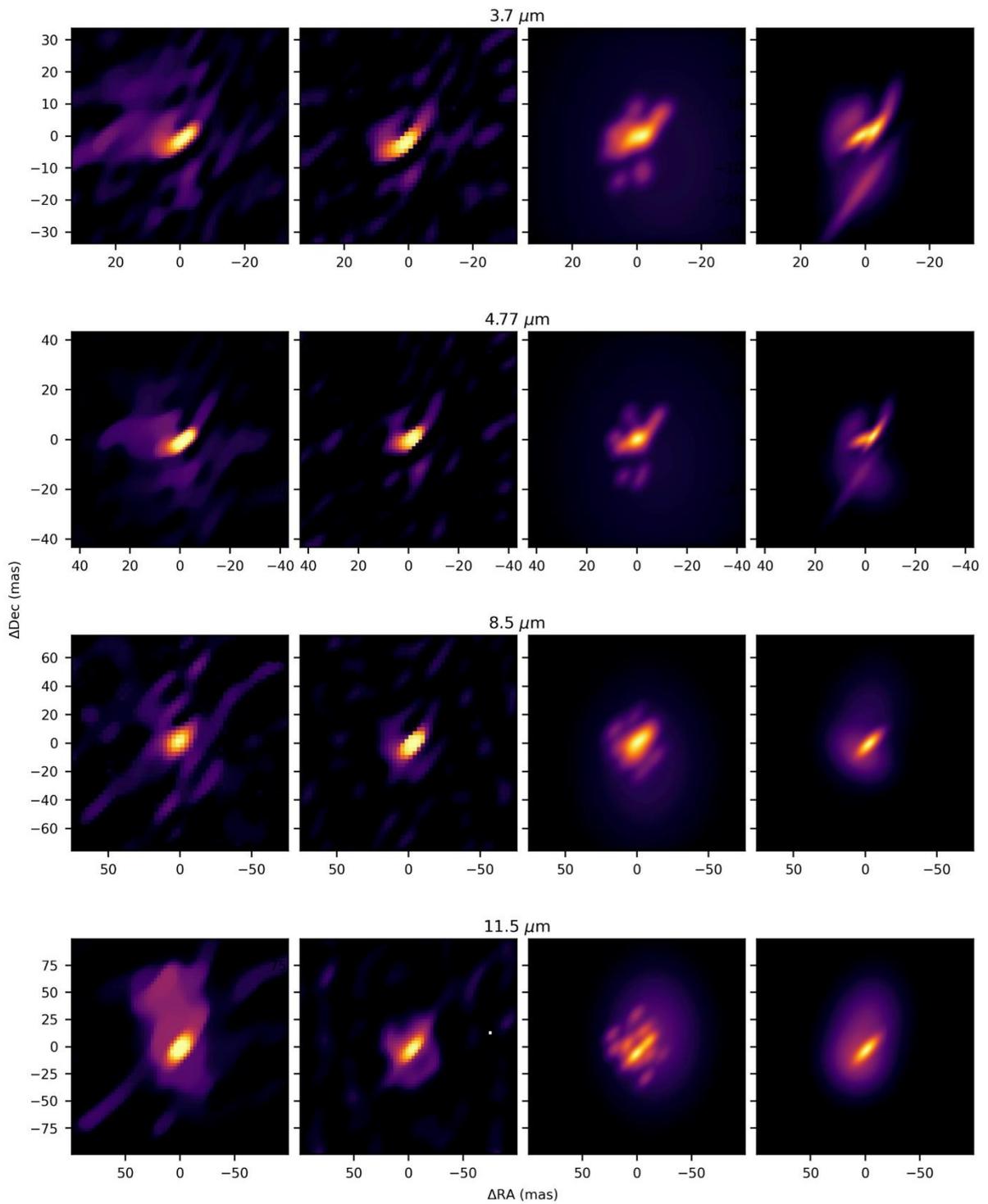

**Extended Data Figure 7: Comparison of reconstructed images at four wavelengths from four algorithms.** *From left to right: the MIRA image reconstruction, the IRBis image reconstruction, the overfitted point source model (convolved with the beam), and Gaussian model for four selected wavelengths. The plot uses a 0.6 power colour scaling for visual purposes. Each method reveals similar structures and morphology.*

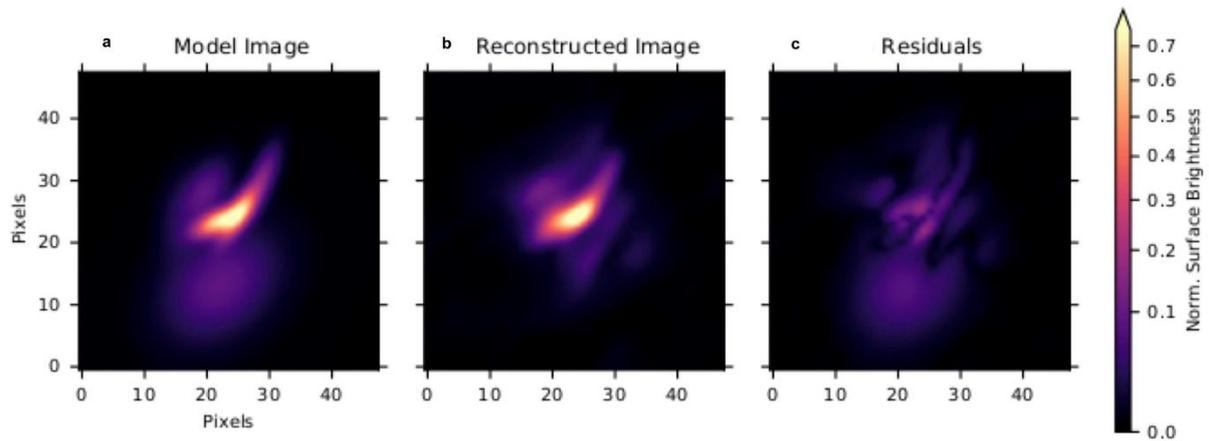

**Extended Data Figure 8: Evaluation of artefacts created by IRBis image reconstruction**: In order to quantify the fidelity of the reconstructions shown in Main Figure 1, we performed analogous reconstruction on an artificial model. The model consisted of seven Gaussians, similar to our multi-Gaussian model for the dust emission (Methods §6.5). We simulated visibility and closure phase data for this model for our uv-coverage; we added noise to the simulated data similar to that in the observations. We then performed image reconstruction using IRBis with identical reconstruction parameters to those used in Main Figure 1. The above Figures represent **a**: The input 7-gaussian model; **b**: The IRBis reconstructed image; **c**: The reconstructed image minus the input model. In all cases the colour scale represents the fraction of the peak intensity of the original model. The r.m.s. errors in the residual maps were 2.3% of the peak brightness. This indicated that most of the artefacts present in Main Figure 1 result from the uv-coverage rather than noise on the observed quantities. In Figure 1 we have drawn white contours at 3 $\sigma$ = 6% of the peak. Features brighter than this certainly represent true source emission.

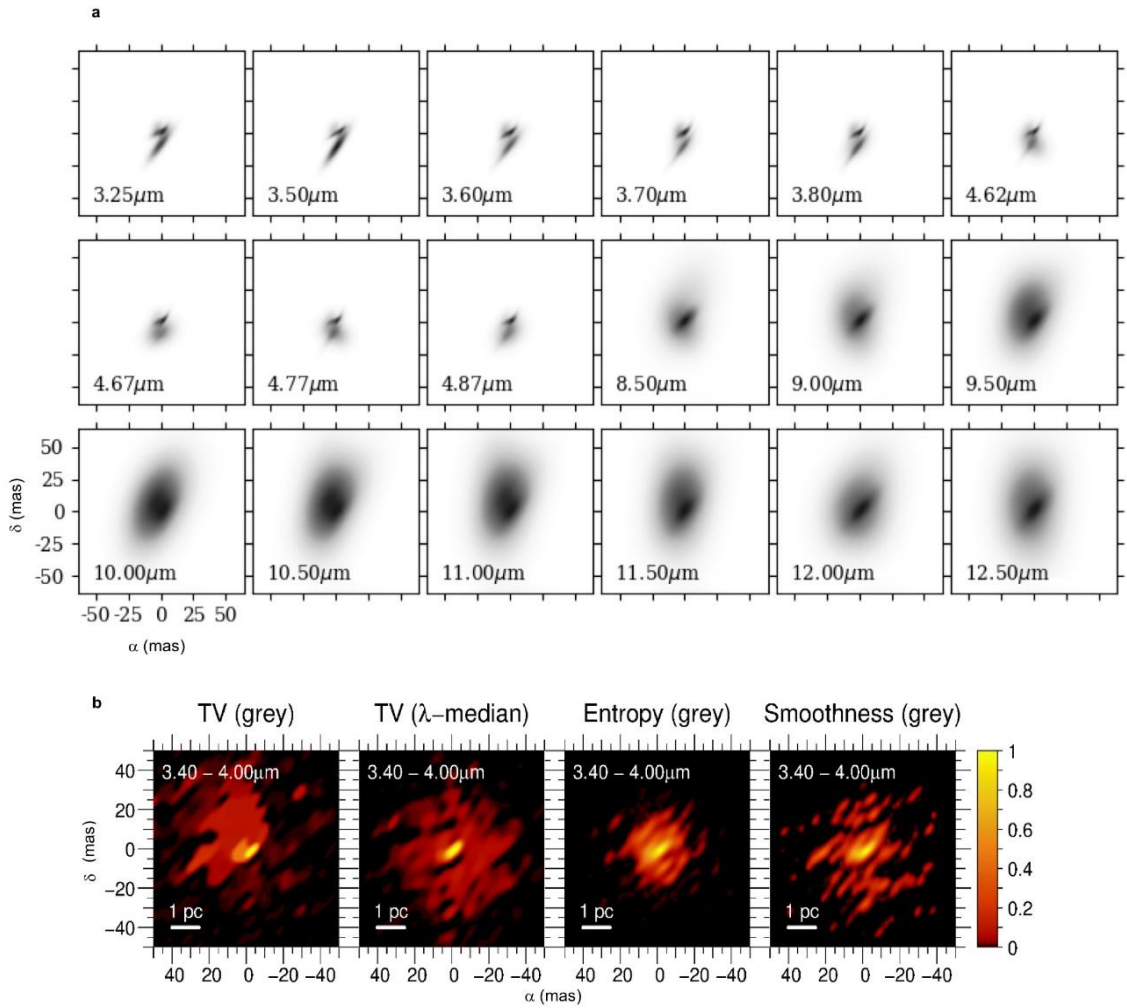

**Extended Figure 9: NGC 1068 image reconstruction data. a:** *Image representation of models obtained from the Gaussian modelling approach described in Methods §6.5. We use square root intensity scale.* **b:** *Image reconstruction with MIRA using different methods c.f. Methods §6.4. From left to right: using Total Variation (TV) regulariser, on a large bandwidth ("grey" reconstruction); using the same regulariser, but independently reconstructing images at each wavelength, and computing a median over the wavelength interval; using Maximum Entropy regulariser (grey reconstruction); using Smoothness regulariser (grey reconstruction).*

### a) IRBis reconstruction parameters

| Wavelength range [μm] | Closure Phase error [deg] (1) | $V^2$ S/R (2) | FOV [mas] (3) | $N_{pix}$ [pixels] (4) | Cost func. (5) | Regularisation (6) | Engine number (7) | uv-weight (8) |
|---|---|---|---|---|---|---|---|---|
| 3.05 - 3.4 | 15,1 | 8,6 | 340 | 256 | 1 | + 2 | 2 | 1 |
| 3.4 - 4.0 | 4 | 13,3 | 340 | 256 | 1 | - 4 | 2 | 0,5 |
| 4.55 - 4.91 | 4,5 | 14,6 | 340 | 256 | 1 | - 4 | 2 | 1 |
| 8.0 - 9.0 | 1,2 | 13,2 | 360 | 128 | 1 | - 4 | 1 | 0,5 |
| 10.0 - 11.0 | 4,3 | 8 | 360 | 128 | 1 | + 2 | 1 | 1 |
| 11.5 - 12.5 | 2,4 | 11,6 | 360 | 128 | 1 | + 2 | 1 | 1 |

### b) L and M bands best fit parameters

| Gauss # | Param. | 3.7 μm | | | | 4.67 μm | | |
|---|---|---|---|---|---|---|---|---|
| 1 | $\Theta$ | 15,8 | + | 4,0 | - 4,0 | 17,7 | + 3,5 | - 4,0 |
| 1 | $\theta$ | 9,8 | + | 2,9 | - 2,5 | 12,6 | + 2,1 | - 2,0 |
| 1 | PA | -38,1 | + | 102,7 | - 38,0 | -57,6 | + 114,7 | - 25,3 |
| 1 | frel | 3,4 | + | 2,5 | - 1,6 | 5,9 | + 1,5 | - 1,6 |
| 1 | l | -0,6 | + | 0,6 | - 0,5 | -1,0 | + 1,0 | - 0,9 |
| 1 | m | 0,9 | + | 0,1 | - 0,1 | 0,9 | + 0,2 | - 0,1 |
| 2 | $\Theta$ | 8,1 | + | 2,0 | - 2,3 | 7,9 | + 1,6 | - 2,5 |
| 2 | $\theta$ | 2,7 | + | 0,8 | - 0,9 | 3,0 | + 1,1 | - 1,5 |
| 2 | PA | -51,8 | + | 151,5 | - 7,8 | -56,9 | + 154,7 | - 13,2 |
| 2 | frel | 4,2 | + | 1,5 | - 1,1 | 4,6 | + 1,2 | - 1,2 |
| 2 | l | -1,5 | + | 1,1 | - 1,0 | -1,3 | + 0,9 | - 0,7 |
| 2 | m | 14,9 | + | 0,6 | - 0,7 | 14,7 | + 0,6 | - 0,6 |
| 3 | $\Theta$ | 13,6 | + | 1,3 | - 1,1 | 13,9 | + 1,6 | - 1,4 |
| 3 | $\theta$ | 5,8 | + | 1,1 | - 1,1 | 6,3 | + 1,9 | - 2,0 |
| 3 | PA | -7,2 | + | 1,2 | - 1,0 | -7,0 | + 1,1 | - 1,3 |
| 3 | frel | 1,6 | + | 1,5 | - 1,1 | 1,0 | + 1,1 | - 0,8 |
| 3 | l | -9,0 | + | 0,8 | - 1,1 | -8,9 | + 1,8 | - 1,3 |
| 3 | m | 15,4 | + | 0,8 | - 1,0 | 15,3 | + 1,4 | - 1,0 |
| 4 | $\Theta$ | 8,9 | + | 1,0 | - 1,0 | 9,2 | + 1,2 | - 1,3 |
| 4 | $\theta$ | 2,6 | + | 0,7 | - 0,8 | 2,2 | + 0,7 | - 0,6 |
| 4 | PA | -32,6 | + | 1,1 | - 1,1 | -33,1 | + 1,5 | - 1,4 |
| 4 | frel | 3,7 | + | 1,5 | - 1,3 | 4,7 | + 1,2 | - 1,1 |
| 4 | l | 3,1 | + | 0,7 | - 0,8 | 3,5 | + 0,7 | - 0,6 |
| 4 | m | 16,6 | + | 1,0 | - 0,8 | 16,4 | + 1,0 | - 0,6 |
| 5 | $\Theta$ | 9,7 | + | 1,2 | - 1,1 | 10,2 | + 1,4 | - 1,2 |
| 5 | $\theta$ | 5,6 | + | 0,9 | - 1,1 | 5,6 | + 1,5 | - 1,4 |
| 5 | PA | -29,9 | + | 1,2 | - 1,1 | -29,9 | + 1,3 | - 1,3 |
| 5 | frel | 2,3 | + | 1,5 | - 1,0 | 1,6 | + 0,8 | - 0,6 |
| 5 | l | -4,0 | + | 0,9 | - 0,8 | -3,6 | + 0,8 | - 0,7 |
| 5 | m | 20,0 | + | 0,8 | - 0,9 | 20,4 | + 1,5 | - 1,7 |
| 6 | $\Theta$ | 9,2 | + | 0,9 | - 1,0 | 9,2 | + 1,0 | - 0,9 |
| 6 | $\theta$ | 3,4 | + | 1,0 | - 1,1 | 3,1 | + 1,0 | - 1,1 |
| 6 | PA | -20,7 | + | 1,1 | - 1,4 | -20,3 | + 1,6 | - 1,7 |
| 6 | frel | 1,0 | + | 1,3 | - 0,7 | 0,7 | + 0,7 | - 0,4 |
| 6 | l | 7,3 | + | 1,1 | - 1,0 | 7,8 | + 1,2 | - 1,3 |
| 6 | m | 25,0 | + | 0,9 | - 1,0 | 24,9 | + 1,6 | - 1,3 |
| 7 | $\Theta$ | 22,5 | + | 1,0 | - 1,0 | 22,5 | + 1,6 | - 1,2 |
| 7 | $\theta$ | 3,2 | + | 1,0 | - 0,9 | 2,8 | + 1,5 | - 1,3 |
| 7 | PA | -35,4 | + | 1,3 | - 1,5 | -35,5 | + 1,7 | - 2,1 |
| 7 | frel | 2,5 | + | 2,0 | - 1,6 | 0,6 | + 1,9 | - 0,5 |
| 7 | l | -2,0 | + | 0,9 | - 1,0 | -1,8 | + 1,3 | - 2,0 |
| 7 | m | -2,7 | + | 1,3 | - 1,3 | -3,2 | + 2,8 | - 1,8 |

### c) N band best fit parameters

| Gauss # | Param. | 8.5 μm | | | | 12 μm | | |
|---|---|---|---|---|---|---|---|---|
| 1 | $\Theta$ | 50,4 | + | 4,9 | - 4,3 | 55,9 | + 4,4 | - 4,6 |
| 1 | $\theta$ | 32,0 | + | 5,1 | - 5,3 | 37,0 | + 3,9 | - 3,7 |
| 1 | PA | -17,1 | + | 14,0 | - 15,6 | -14,5 | + 12,7 | - 11,5 |
| 1 | frel | 1,0 | + | 0,0 | - 0,0 | 1,0 | + 0,0 | - 0,0 |
| 1 | l | 0,0 | + | 0,0 | - 0,0 | 0,0 | + 0,0 | - 0,0 |
| 1 | m | 0,0 | + | 0,0 | - 0,0 | 0,0 | + 0,0 | - 0,0 |
| 2 | $\Theta$ | 15,8 | + | 1,2 | - 1,3 | 22,8 | + 3,0 | - 3,4 |
| 2 | $\theta$ | 5,9 | + | 0,7 | - 0,9 | 8,0 | + 1,4 | - 1,5 |
| 2 | PA | -43,8 | + | 3,0 | - 2,7 | -42,1 | + 7,5 | - 5,5 |
| 2 | frel | 1,2 | + | 0,4 | - 0,3 | 0,4 | + 0,7 | - 0,1 |
| 2 | l | -3,2 | + | 2,6 | - 2,4 | -2,0 | + 2,6 | - 2,5 |
| 2 | m | 15,0 | + | 1,9 | - 1,8 | 17,1 | + 2,0 | - 1,8 |
| 3 | $\Theta$ | 26,7 | + | 2,8 | - 3,1 | 34,5 | + 16,0 | - 3,4 |
| 3 | $\theta$ | 19,7 | + | 2,6 | - 3,2 | 32,6 | + 3,1 | - 4,1 |
| 3 | PA | 50,6 | + | 14,0 | - 23,3 | 6,2 | + 47,7 | - 59,6 |
| 3 | frel | 1,4 | + | 0,7 | - 0,5 | 0,7 | + 1,0 | - 0,2 |
| 3 | l | 3,2 | + | 2,7 | - 2,7 | 7,3 | + 3,6 | - 6,2 |
| 3 | m | 21,6 | + | 2,2 | - 2,4 | 26,5 | + 2,8 | - 13,5 |

### d) SED parameters

| Aperture | Carbon content % | $T_{cold}$ [K] min .. max | $\eta_{cold}$ min .. max | $T_{hot}$ [K] min .. max | $\eta_{hot}$ min .. max | $N_{ext}$ [μg/cm²] min .. max | $\chi^2$ min .. max |
|---|---|---|---|---|---|---|---|
| E1 | 0 | 561 .. 581 | 0,04 .. 0,06 | N.A. | N.A. | 396 .. 547 | 10 .. 13 |
| E1 | 20 | 666 .. 701 | 0,04 .. 0,04 | N.A. | N.A. | 448 .. 610 | 15 .. 18 |
| E2 | 0 | 188 .. 469 | 0,04 .. 1,00 | 703 .. 1179 | 0,00050 .. 0,00050 | 281 .. 530 | 7 .. 10 |
| E2 | 20 | 206 .. 506 | 0,04 .. 0,62 | 829 .. 1461 | 0,00100 .. 0,00660 | 386 .. 728 | 8 .. 11 |
| E3 | 0 | 217 .. 376 | 0,04 .. 0,50 | 860 .. 1095 | 0,00050 .. 0,00130 | 67 .. 386 | 8 .. 11 |
| E3 | 20 | 225 .. 371 | 0,05 .. 0,34 | 917 .. 1412 | 0,00050 .. 0,00130 | 92 .. 489 | 8 .. 12 |
| E4 | 0 | 232 .. 389 | 0,06 .. 0,43 | 555 .. 818 | 0,00060 .. 0,00950 | 281 .. 530 | 4 .. 7 |
| E4 | 20 | 250 .. 412 | 0,09 .. 0,38 | 633 .. 1200 | 0,00040 .. 0,00940 | 386 .. 621 | 4 .. 7 |
| E5 | 0 | 223 .. 285 | 0,10 .. 0,36 | 717 .. 1098 | 0,00005 .. 0,00040 | 1 .. 100 | 6 .. 9 |
| E5 | 20 | 220 .. 287 | 0,04 .. 0,38 | 712 .. 1275 | 0,00004 .. 0,00040 | 1 .. 189 | 6 .. 10 |
| DE1 | 0 | 229 .. 238 | 0,10 .. 0,13 | 771 .. 865 | 0,00001 .. 0,00002 | 1 .. 34 | 41 .. 45 |
| DE1 | 20 | 228 .. 242 | 0,09 .. 0,13 | 820 .. 870 | 0,00001 .. 0,00001 | 1 .. 43 | 41 .. 45 |
| DE2 | 0 | 139 .. 228 | 0,09 .. 1,00 | 417 .. 475 | 0,00299 .. 0,00918 | 825 .. 1000 | 12 .. 16 |
| DE2 | 20 | 153 .. 292 | 0,06 .. 1,00 | 543 .. 703 | 0,00175 .. 0,00775 | 1000 .. 1468 | 11 .. 14 |
| DE3 | 0 | 148 .. 162 | 0,40 .. 1,00 | N.A. | N.A. | 1 .. 197 | 13 .. 17 |
| DE3 | 20 | 148 .. 162 | 0,40 .. 1,00 | N.A. | N.A. | 1 .. 241 | 13 .. 17 |
| SE | 0 | 203 .. 354 | 0,03 .. 0,51 | 846 .. 1031 | 0,00060 .. 0,00140 | 530 .. 853 | 14 .. 18 |
| SE | 20 | 173 .. 300 | 0,04 .. 1,00 | 1333 .. 1500 | 0,00070 .. 0,00090 | 464 .. 619 | 27 .. 31 |

*Extended Data Table 1: Parameters for IRBis image reconstruction, Gaussian modelling, and SED fitting. a) Parameters used in the IRBis image reconstructions. 1) Mean closure phase error in the data (deg); 2) Mean signal-to-noise ratio of the squared visibility in the data; 3) Field of View of the reconstruction in mas; 4) Pixel grid dimensions (x=y) used in the reconstruction; 5) Cost function number as defined in the text; 6) Regularisation functions numbers: 2 and 4 correspond respectively to maximum entropy and total variation and signs indicate used prior: + means a fitted Gaussian and – means a constant; 7) Optimisation engine number as defined in the text; 8) Power of the uv-density weight. b) L and M bands best fit parameters. $\Theta$: FWHM of the major axis in milliarcseconds (mas), $\theta$: FWHM of the minor axis (mas), PA: position angle of the major axis (degrees), frel: relative integrated flux of the gaussian component (%), l : position west of the centre for the component (mas), m : position north of the centre (mas). c) N band best fit parameters. Symbols as in b. d) Extended list of SED parameters, for each area defined in figure 2, assuming either 0% or 20% carbon content (by weight) in the absorbing screen. $T_{hot}$ and $T_{cold}$ are temperatures and $\eta_{hot}$ and $\eta_{cold}$ are the filling factors of the 2*

blackbodies, except for E1 and DE3 that are adequately described by a single cold blackbody. $N_{ext}$ is the extinction and $\chi^2$ refers to the fit quality.